# Ultra-narrow linewidth light generation based on an optoelectronic oscillator

Qizhuang Cen[1,2,3,†], Shanhong Guan[1,2,3,†], Dongdong Jiao[4,5], Tengfei Hao[1,2,3], X. Steve Yao[6,7], Yitang Dai[8,*], Ming Li[1,2,3,*]

[1]Key Laboratory of Optoelectronic Materials and Devices, Institute of Semiconductors, Chinese Academy of Sciences; Beijing 100083, China.

[2]College of Materials Science and Opto-Electronic Technology, University of Chinese Academy of Sciences, Beijing 100049, China.

[3]School of Electronic, Electrical and Communication Engineering, University of Chinese Academy of Sciences, Beijing 100049, China

[4]National Time Service Center, Chinese Academy of Sciences; Xi'an 710600, China.

[5]Key Laboratory of Time and Frequency Standards, Chinese Academy of Sciences; Xi'an 710600, China.

[6]Photonics Information Innovation Center and Hebei Provincial Center for Optical Sensing Innovations, College of Physics Science & Technology, Hebei University; Baoding 071002, China.

[7]NuVision Photonics, Inc.; Las Vegas, NV 89109, USA.

[8]State Key Laboratory of Information Photonics and Optical Communications, Beijing University of Posts and Telecommunications; Beijing 100876, China.

†These authors contribute equally.

*Corresponding authors. Email: ytdai@bupt.edu.cn; ml@semi.ac.cn;

**Abstract:** Narrow-linewidth light sources are essential for both fundamental research and various technological applications, yet they are challenging to generate directly due to instabilities in active laser cavities and the misalignment between closely spaced resonator modes and broad gain bandwidth. In this study, we demonstrate the direct generation of low-noise light from an optoelectronic oscillator (OEO). By minimizing the delay in the optoelectronic link and employing a high-quality optical resonator, an OEO can simultaneously generate both a low-phase-noise optical oscillation that is resilient to phase fluctuations of its pump laser, and a radio frequency (RF) oscillation that captures these fluctuations. We leverage these RF-domain fluctuations to implement a high-performance feedback loop, which stabilizes the pump laser and significantly enhances the performance of optical oscillation. This approach achieves an outstanding phase noise level of −100 dBc/Hz at a 1 kHz offset and an integrated linewidth as narrow as 0.23 Hz. The integration of optoelectronic oscillation and feedback loop provides significant broadband noise suppression compared to conventional schemes for generating narrow-linewidth light, paving the way for developments in fields such as coherent optical communications, atomic spectroscopy, metrology, and quantum optics.

## Introduction

Narrow-linewidth and low-noise light sources are pivotal in advancing various fields, including coherent optical communications (*1*), atomic spectroscopy (*2*), metrology (*3-6*), and quantum optics (*7*). Traditionally, high-purity light is generated within a laser cavity that houses the pump and gain medium, where gain and loss are balanced (*8*). The spectral purity of a laser is fundamentally determined by the cavity's quality factor (Q-factor) and its stability in the physical dimension. Nowadays, high-$Q$ optical resonators such as Fabry–Pérot (FP) cavities (*9, 10*), crystalline whispering gallery mode (WGM) resonators (*11, 12*), and integrated microresonators (*13, 14*), are increasingly prevalent. Ultra-stable resonators are also available through temperature stabilization, the use of low thermal expansion material, and vibration isolation (*9, 15*). However, directly generating low-noise, narrow-linewidth light from these high-Q resonators presents challenges due to mode competition and hopping caused by the mismatch between the closely spaced resonator mode (GHz) and the large optical gain bandwidth (THz) (*8, 16*).

An effective strategy to harness high-Q resonators is to lock a moderate-performance laser to the target resonator through self-injection locking technology (*11, 12, 17-20*). In this process, the light oscillates between the active laser cavity and the external passive high-Q resonator, significantly suppressing phase noise and compressing the linewidth. Self-injection locking has enabled lasers to achieve fundamental linewidths as low as sub-100 mHz (*17*) and integral linewidths below 100 Hz (*19, 20*). Theoretically, the suppression factor of the phase noise is approximately proportional to the ratio of the Q-factors between the external cavity and the laser cavity across a broad frequency offset range. However, gain nonlinearities and temperature fluctuations within the active laser cavity hinder the effective suppression, especially at low-frequency offsets, thus preventing the output from reaching the frequency stability limit set by the external passive high-Q resonator (*21, 22*). Active frequency-locking schemes using a feedback loop can effectively transfer the stability of an external passive resonator to a laser. Among these schemes, the Pound–Drever–Hall (PDH) technique stands out as the most popular and efficient scheme, enabling the development of advanced lasers with integral linewidths below 10 mHz (*23-25*). Despite its effectiveness, the noise suppression capabilities of PDH and other feedback-based techniques are limited by the bandwidth of the feedback loop, and suppression effectiveness decreases with increasing frequency offsets. Additionally, the feedback control mechanism introduces servo bumps, posing further challenges.

The synergistic integration of oscillation and feedback mechanisms offers the potential for significant broadband noise suppression. Previous studies have successfully incorporated this integration into a phase-modulation-to-intensity-modulation (PM-IM) optoelectronic oscillator (OEO) system to generate narrow-linewidth light (*26-30*). However, in this demonstration, locking two lasers to adjacent longitudinal modes of the same cavity led to common-mode rejection of the cavity's thermal noise. This issue hindered a comprehensive performance evaluation and ultimately restricted the widespread adoption of this technique for generating narrow-linewidth light.

In this work, we demonstrate that with the integration of optoelectronic oscillation and feedback loop, a PM-IM OEO system can generate low-noise, narrow-linewidth light, with performance approaching the thermal noise limit of an ultra-stable optical resonator. Both our theoretical analyses and experimental results show that the PM-IM OEO with a high-Q optical resonator and a shortened optoelectronic link acts as an optoelectronic frequency converter, transforming the pump laser (PL) into an RF signal. This downconversion not only enables the generation of low-noise optical oscillation but also facilitates a high-performance feedback loop to stabilize the PL. Our results reveal that the OEO-locking scheme provides extensive and wide-ranging noise suppression capabilities thanks to the integration of optoelectronic oscillation and feedback loop, positioning it as a superior alternative to the conventional Pound-Drever-Hall (PDH) locking scheme.

## Principle

The generation of low-noise light based on a PM-IM OEO is illustrated in Fig.1A. This system primarily includes a passive high-Q optical resonator—specifically, an FP cavity in our demonstration—that stores low-noise optical oscillation; an active optoelectronic link that provides gain and supports RF oscillation; a PL used as the optical carrier of the OEO; and a feedback module that stabilizes the PL. The optical oscillation within the resonator comes from the injection of one of the first-order sidebands of the phase-modulated light. The angular frequencies of the oscillations are related by $\omega_{RF} = |\omega_{PL} - \omega_O|$, where $\omega_{RF}$, $\omega_{PL}$, and $\omega_O$ are the angular frequencies of the RF oscillation, the PL, and the optical oscillation, respectively. The phase fluctuations of the PL are redistributed between the RF and the optical sidebands during the oscillation process. When using a stable, high-Q resonator and minimizing the delay in the optoelectronic link, these phase fluctuations are almost transferred to the RF and hardly ever to the optical oscillation, as shown in Fig. 1B. By comparing the OEO-generated RF to a stable reference, the phase fluctuations are extracted and used to stabilize the PL, further eliminating its impact on the optical oscillation, as shown in Figs. 1C and 1D.

The self-alignment mechanism in the PM-IM OEO leads to the redistribution of the PL's phase noise between the optical and RF oscillations. The total reflected light from the FP cavity is the coherent sum of the promptly reflected light and the cavity's leakage light, as illustrated in Fig. 1B (*23, 25*). The promptly reflected light, which is still purely phase-modulated, generates only the direct current (DC) component at the photodetector (PD). The RF signal is generated solely from the beat note between the optical carrier and the leakage light. Effective PM-IM conversion occurs only when one of the sidebands (assuming the lower sideband) aligns with an FP cavity resonance. This selective PM-IM conversion forms an equivalent RF bandpass filter (BPF) with the same filter window as that of the FP cavity and center frequency equal to the frequency difference between the PL and the FP cavity resonance, as illustrated in Fig. 1E (*31*) (supplementary information section I). Consequently, the RF frequency component within the optoelectronic cavity that matches this RF BPF is selectively amplified after each roundtrip, leading to a stable oscillation. When the frequency of the PL drifts, the center frequency of the RF BPF adjusts accordingly. Based on the minimum loss principle, the RF frequency follows this drift, and the optical oscillation automatically aligns with the cavity resonance. This self-adaptive dynamic allows the PM-IM OEO to continuously generate low-noise light. Thanks to the narrow linewidth of the FP cavity and manageable gain bandwidth in the optoelectronic link, single-frequency operations in both optical and RF domains are guaranteed.

Ideally, the optical oscillation would align precisely with the FP cavity resonance. However, this alignment can be compromised by the phase conditions of the Barkhausen stability criterion in feedback oscillators (*32*). Since the RF is generated from the beat note between the optical carrier and the leakage light, an additional phase shift of the optical oscillation that intruded from the interaction with the FP cavity is imparted to the RF. As a result, the roundtrip phase shift of the RF is expressed as $\omega_{RF}\tau_{RF} + \omega_O\tau_O = 2N\pi$, where $\tau_{RF}$ is the delay of the optoelectronic link excluding the FP cavity and $\tau_O$ is the group delay of the FP cavity at the resonance, respectively (supplementary information sections II and IV). An angular frequency drift of $\Delta\omega_{RF}$ results in an additional phase shift of $\Delta\omega_{RF}\tau_{RF}$. This phase shift needs to be compensated by the phase adjustment in the leakage light to restore the phase conditions for stable oscillation. The Barkhausen phase condition, along with the frequency relationship of $\omega_{RF} = |\omega_{PL} - \omega_O|$, results in the redistribution of the PL's phase fluctuations between the two oscillations, which can be expressed as:

$$\begin{cases} \varphi_O(t) = \dfrac{\tau_{RF}}{\tau_{RF} + \tau_O}\varphi_{PL}(t) \\ \varphi_{RF}(t) = \pm\dfrac{\tau_O}{\tau_{RF} + \tau_O}\varphi_{PL}(t) \end{cases}, \tag{1}$$

where $\varphi_{PL}(t)$ is the phase fluctuations of the PL, and $\tau_{RT} = \tau_{RF} + \tau_O$ is considered as the equivalent roundtrip time of the optoelectronic cavity, the sign "$\pm$" indicates that the lower or upper sideband aligns with the FP cavity response (supplementary information section II). By using a narrow bandwidth FP cavity and minimizing $\tau_{RF}$, we can achieve a condition where $\frac{\tau_O}{\tau_{RF}+\tau_O} \approx 1$ and $\frac{\tau_{RF}}{\tau_{RF}+\tau_O} \approx 0$. This indicates that the majority of the PL's phase fluctuations are effectively transferred to RF oscillation, leaving only a minimal residual in the optical oscillation. Consequently, low-noise light is directly generated from the PM-IM OEO.

Considering that the PL is generally noisy, even if the phase noise of the PL is significantly suppressed, the desired optical oscillation is still degraded by the PL due to the inherent delay of $\tau_{RF}$ in the optoelectronic link. Fortunately, this issue can be eliminated by extracting the phase fluctuations in the RF signal and using them to stabilize the PL. This feedback further suppresses phase noise in the optical oscillations across the feedback bandwidth, as shown in Figs. 1C and 1D. Since the error signal represents the phase fluctuations of the PL, this feedback control is considered as a quasi-phase-locking loop that effectively eliminates tiny frequency fluctuations and provides superior locking accuracy compared to frequency-locking schemes. This dual-suppression mechanism integrating optoelectronic oscillation and feedback loop provides large and broadband noise suppression for the PL, thereby driving the phase noise performance of the optical oscillation to the limit set by the thermal noise of the passive cavity at low-frequency offsets and stochastic intrinsic noise of the active optoelectronic link at high-frequency offsets (supplementary information section II).

## Implementation and Measurement

Two PM-IM OEO systems were built to evaluate the performance of the OEO-locking scheme, as shown in Fig. 2 (see methods). Two external-cavity semiconductor lasers with linewidths of a few kHz were used as the PLs of the OEOs. Two perpendicular FP cavities with free spectral ranges (FSRs) of approximately 3 GHz are positioned horizontally in a cubic vacuum optical cavity. By measuring the equivalent RF BPFs based on the two FP cavities, we obtained 6.26-kHz and 6.15-kHz linewidths of the FP cavities, respectively, corresponding to 50.8-μs and 51.8-μs group delays at the resonance, as shown in Figs. 3A and 3B (supplementary information section I for details). The delays in the optoelectronic link of each OEO are about 10 ns, leading to suppressions of the PL's phase noise by −74.1-dB and −74.3-dB in the desired optical oscillations, respectively. To further suppress the residual phase noise, feedback modules are incorporated into the systems. In the current setup, collecting the low-noise optical oscillation with sufficient power from the cavity transmission poses a challenge due to the low coupling efficiency in the super cavity system. As a workaround, we use the generated RFs to frequency-shift the corresponding PLs, thereby obtaining the low-noise, narrow-linewidth lights that share similar noise performance with the corresponding optical oscillations, except that the noise level is higher at high-frequency offsets. The frequency-shifted PLs without and with feedback are named as the free-running FS-PL and OEO-locked FS-PL, respectively.

The top inset of Figure 3C shows the power spectrum of the RF oscillation from one of the free-running OEO systems. The stable, single-frequency operation differs from conventional long-fiber OEOs, where mode competition and supermode spurs are common. In the max-hold mode, a continuous oscillatory span over 30 MHz is observed, indicating that the system can tolerate such frequency drifts for the PL. Under a constant roundtrip gain, the oscillatory span is approximatively inversely proportional to $\tau_{RF}$. Minimizing the delay in the optoelectronic link will increase the oscillatory span, thereby enhancing the stability (see supplementary materials section IV for details). Similar to conventional OEO, discrete modes with intervals equal $1/\tau_{RF}$ were also observed, as shown in the bottom inset of Fig. 3C. By tuning the PLs, the generated RF oscillations in both OEOs were centered around 400 MHz. These two RFs were used to frequency-shift their respective PLs, obtaining low-noise FS-PLs. The beat note between the FS-PLs is shown in Fig. 3C, further confirming the single-frequency operation of each OEO system.

The phase fluctuations of the PL and its corresponding RF and FS-PL in one of the OEO systems were measured and shown in Fig. 4A. The results prove that the RF accurately replicated the PL's phase fluctuations, and the FS-PL effectively isolated these fluctuations. These findings are further confirmed by the measurement of the frequency fluctuations (Fig. 4B). The frequency fluctuations of FS-PL share the same trend as that of the PL but are suppressed by a factor of about 1/5,000, close to the theoretical prediction of $\tau_{RF}/\tau_{RT}$. A suppression factor of −73.5 dB, close to the theoretical prediction of $20log_{10}\frac{\tau_{RF}}{\tau_{RT}}$, in phase noise spectrum of the free-running FS-PL is also observed, as shown in Fig. 5A. At frequency offsets above 10 kHz, the phase noise of the free-running FS-PL reaches a noise floor set by the amplified intrinsic noise in the active optoelectronic link. By optimizing the optoelectronic cavity, the gain in the RF domain can be significantly reduced, which in turn lowers the noise floor. Ultimately, this reduction can approach half of the relative intensity noise of the PL, which is the same as for conventional OEOs (*28*) (supplementary information section II for details). The linewidth of the free-running FS-PL is measured to be 5 Hz (middle of Fig. 4C), a significant reduction from the 3.5-kHz linewidth of the PL (top of Fig. 4C). To the best of our knowledge, the 5-HZ linewidth is the narrowest value reported to date in free-running optical oscillators. The frequency instability, quantified as modified Allan deviation, is also significantly decreased from $10^{-9}$ level to $10^{-13}$ level, as shown in Fig. 5B. These results indicate that the PM-IM OEO can directly generate low-noise and narrow-linewidth light.

The feedback modules were activated to further suppress the residual phase noise in FS-PL. This feedback confined the frequency fluctuations of the beat signal between two OEO-locked FS-PLs to just a few hertz, as depicted at the bottom of Fig. 4B. The corresponding linewidth is 0.33 Hz, as shown in bottom of Fig. 4C. This suggests an estimated individual linewidth of 0.23 Hz for each OEO-locked FS-PL, assuming equal contributions from each. The phase noise of OEO-locked FS-PL reaches the thermal noise limit set by the thermal noise of the FP cavity at frequency offsets below 1 kHz (Fig. 5A). Above 1-kHz frequency offsets, the intrinsic noise dominates the phase noise. It is important to note that optical oscillation from the transmission port of the FP cavity exhibits significantly lower phase noise compared to the FS-PL at higher frequency offsets due to filtering of the FP cavity. Thanks to the synergy of the optoelectronic oscillation and the feedback loop, we achieved an unprecedented low phase noise of −100 dBc/Hz @ 1kHz. The observable hump around the 10-Hz frequency offset is caused by ambient vibrations. The feedback loop also reduced the frequency instabilities of the FS-PL to $1.73 \times 10^{-15}$ at a 1-s averaging time, which is close to the thermal noise limit of the FP, as shown in Fig. 5B.

The PDH technique is widely recognized as a highly effective method for generating narrow-linewidth light. Here, two PDH systems using the same FP cavities were constructed to compare the performance of PDH scheme against OEO-locking scheme. As shown in Fig. 5C, the OEO-locked PL achieves lower phase noise than that measured in the PDH systems. With further frequency-shifting, the OEO-locked FS-PL shows a significant improvement over the PDH scheme, with reductions of 26 dB and 39 dB at 1 kHz and 10 kHz frequency offsets, respectively. The simulation comparison provided in section V of the supplementary information also supports the experimental results. The improvement is also confirmed by the frequency stability measurements, as illustrated in Fig. 5D. The OEO-locking scheme shows reduced frequency instability across averaging times ranging from 0.1 to 100 seconds, compared to that of the PDH system.

## Discussion and outlook

Thanks to its great self-alignment capability, the PM-IM OEO can directly generate low-noise light without a feedback loop, proving particularly advantageous in scenarios where high-performance feedback loops are impractical. Minimizing the delay in the optoelectronic link through the use of high-speed, integrated devices and microassembly techniques could significantly improve noise suppression capabilities (*33, 34*). With optimal design, the oscillation mechanism alone can achieve a suppression as large as $20log_{10}(F)$, where $F$ denotes the cavity's finesse. This results in a suppression of over 120 dB when using a million-finesse cavity (*10*) (supplementary information section III). When combined with a low-drift or pre-stabilized PL, the phase noise of the generated light oscillation can reach the thermal noise limit of an ultra-stable passive optical resonator, enabling the direct generation of light with sub-Hz integrated linewidth.

Compared to conventional laser where light generation is generally based on stimulated emission of electromagnetic radiation, the low-noise light generation in the PM-IM OEO is accomplished with a PL and a RF LAN incorporated with optoelectronic interconversion. This novel gain regime is characterized by small nonlinearity, insensitivity to temperature, and easy-to-manage gain bandwidth, resulting in the direct generation of single-mode, low-noise light. Due to the deliberate shortening of the optoelectronic link and its simple configuration, the PM-IM OEO is well-suited for further integration. As integrated high-Q optical resonators (*10, 13, 14*) and lasers (*35, 36*) become increasingly available, they can be seamlessly integrated into the PM-IM OEO, offering a compact, low-power, and cost-effective low-noise light source. Moreover, given that the PM-IM OEO functions as a voltage-controlled RF oscillator, commercial chip-scale phase-locked loop (PLL) circuits can be integrated into the PM-IM OEO system (*37*), eliminating any residual phase noise of the PL's noise in the optical oscillation. The collective integration of high-Q resonators,

PLL circuits, and the optoelectronic link enables miniaturization of the PM-IM OEOs. The low-noise lights generated from miniature OEOs can reach the phase noise floor set by the thermal noise of the high-Q optical resonators. Thanks to the novel gain regime, the PM-IM OEO has the potential to outperform existing self-injection-locking laser systems, where gain nonlinearities still severely destabilize output frequency.

With the incorporation of a feedback loop, the PM-IM OEO shares structural similarities with the PDH scheme. The similarity facilitates the adoption of our scheme from the existing PDH framework, enabling a transition to a more robust, easily implemented, and high-performance ultra-narrow linewidth system. Compared to PDH systems and self-injection-locking lasers, which rely only on feedback or oscillation mechanisms to suppress the noise, our approach integrates with the oscillation and feedback mechanisms, representing a significant shift in the approach to generating low-noise, narrow-linewidth light. This synergy decreases dependency on the PL's noise performance and the Q-factor of the optical resonator. Compared with the mature PDH and self-injection-locking schemes that have been in development for decades, the OEO-locking approach is primitive. It presents significant potential for optimization and enhancement, promising substantial benefits for coherent optical communication, precision metrology, and quantum optics applications.

## Methods

**FP cavity fabrication:** Two perpendicular FP cavities are positioned horizontally in a cubic optical cavity with a length of 50 mm. The eight vertices of the cubic cavity are chamfered towards the center, each with a chamfer depth of approximately 4.2 mm. Both cavities' spacers and mirror substrates are constructed using standard-grade, ultra-low expansion (ULE) glass. The two mirrors have a diameter of 25.4 mm, and a thickness of 6.3 mm, and are coated with a high-reflectivity layer for 1550 nm light. One mirror is flat (infinite radius of curvature), while the other has a curvature radius of 500 mm. The linewidths of the optical cavities are measured to be 6.26 kHz and 6.15 kHz in two directions, corresponding to finesse values of approximately 479,200 and 487,800, respectively. To support the optical cavity, a thermal shield, a bracket, and a vacuum flange are interconnected using four stainless steel screws, which are the same as in reference (*38*). Polyetheretherketone (PEEK) columns are installed between the bracket and the thermal shield to prevent heat transfer. The cavity is affixed to the bracket with four PEEK screws, exerting a squeezing force of around 100 $N$ at each vertex. To mitigate external factors such as temperature fluctuations, and to reduce airborne sound, the cavity, and its support system are enclosed within a vacuum chamber, maintained at a pressure of less than approximately $5 \times 10^{-6}$ Pa.

**Experimental setup:** The experimental setup, as depicted in Fig. 2, utilized two commercial external-cavity semiconductor lasers (LXNLM, Smartcore) operating around 1550.92 nm as the PLs, with a frequency difference of approximately 1.8 GHz. By adjusting the voltage applied to the PZT, each laser can achieve a frequency shift of up to ± 600 MHz. In each OEO system, a 20-mW output beam was split into two branches by a 3-dB polarization-maintaining fiber coupler. One branch served as the optical carrier for the OEO, while the other was frequency-shifted through an AOM by the OEO-generated RF. This frequency-shifting generated low-phase-noise light, which was then utilized for further evaluation and potential applications. Before entering the PM, the optical power was attenuated to 1.5 mW to minimize the effects of power fluctuations on the FP cavity. After the PM, the phase-modulated light was directed into the FP cavity. The reflected light from the FP cavity was then directed to a PD with a bandwidth of 600 MHz. The RF signal generated in the PD was amplified by an LNA and subsequently split into three branches by two RF power splitters. One branch was fed back to the PM, while the other two branches were directed to the AOM and the feedback module, respectively. The AOM operates at a central frequency of 400 MHz with an operational bandwidth of 40 MHz. Within the feedback module, the RF signal was frequency-divided and mixed with an external reference generated from the arbitrary waveform generator (AWG). The mixer output passed through a low-pass filter (LPF) with a bandwidth of 1 MHz before being fed into a proportional-integral (PI) controller (New Focus, LB1005). The output from the PI controller was then routed back to the PZT tuning port of the PL. When the feedback control system was active, the frequency-divided RF was phase-locked to the reference, ensuring that the optical oscillation remained in resonance with the FP cavity.

**Data measurement:**

Power spectrum measurement: The power spectra shown in Fig. 3C were measured using a power spectrum analyzer (FSW, Rohde & Schwarz). The oscillatory frequency range was measured in the max-hold mode by tuning the PL frequency.

Phase fluctuations measurement: The free-running RF was recorded with a real-time oscilloscope (RTO2004, Rohde & Schwarz) with a sampling rate of 2.5 GS/s in Fig. 4A. The corresponding phase fluctuations were demodulated using a Matlab algorithm. With the same method, the phase fluctuations of the free-running PL and FS-PL were also obtained by measuring the beat signal between the free-running PL and an OEO-locked FS-PL, assuming the phase fluctuations of the OEO-locked FS-PL are negligible. These data were recorded simultaneously.

Frequency stability measurement: The frequency fluctuations in Fig. 4B were recorded by frequency counters (Keysight, 53230A). Specifically, the free-running PL (top of Fig. 4B) and the FS-PL (middle of Fig. 4B) were downconverted to the RF domain with an OEO-locked FS-PL, assuming the contribution from OEO-locked FS-PL is negligible. Note that in the free-running OEO, the frequency fluctuations of the PL, the FS-PL, and the RF were simultaneously recorded. The relative frequency fluctuations between two OEO-locked FS-PL, shown in Fig. 4B (bottom), were similarly processed and recorded. These recorded data were used to calculate the modified Allen deviation presented in Fig. 5. The modified Allen deviation of the PDH scheme in Fig. 5D was also obtained by measuring the frequency fluctuations of the beat note between the two PDH systems.

Linewidth measurement: The linewidth of the free-running PL shown in Fig. 4C (top) was obtained by measuring the beat signal between a free-running PL and an OEO-locked FS-PL using a power spectrum analyzer (FSW, Rohde & Schwarz) under the RWB of 1 kHz. The linewidth of the free-running FS-PL shown in Fig. 4C (middle) was obtained by measuring the beat signal between a free-running FS-PL and an OEO-locked FS-PL using a fast Fourier transform (FFT) analyzer (Keysight, 35670A). The RBW for this measurement was set to be 2 Hz, corresponding to the measurement duration of 0.5 seconds. The linewidth of the OEO-locked FS-PL shown in Fig. 4C (bottom) was obtained by measuring the beat signal between two OEO-locked FS-PLs using the FFT analyzer. The RBW for this measurement was set to be 0.125 Hz, corresponding to the measurement duration of 8 seconds.

Phase noise spectrum measurement: The phase noise spectra in Fig. 5 were measured using a phase noise analyzer (FSWP, Rohde & Schwarz). The phase noise spectrum of the free-running PL was obtained by measuring the beat signal between a free-running PL and an OEO-locked FS-PL, assuming the contribution from the OEO-locked FS-PL is negligible. The phase noise spectra of OEO-locked PL/FS-PL and PDH-locked laser were obtained by measuring the beat signal between two OEO-locked PL/FS-PL and PDH-locked lasers, respectively.

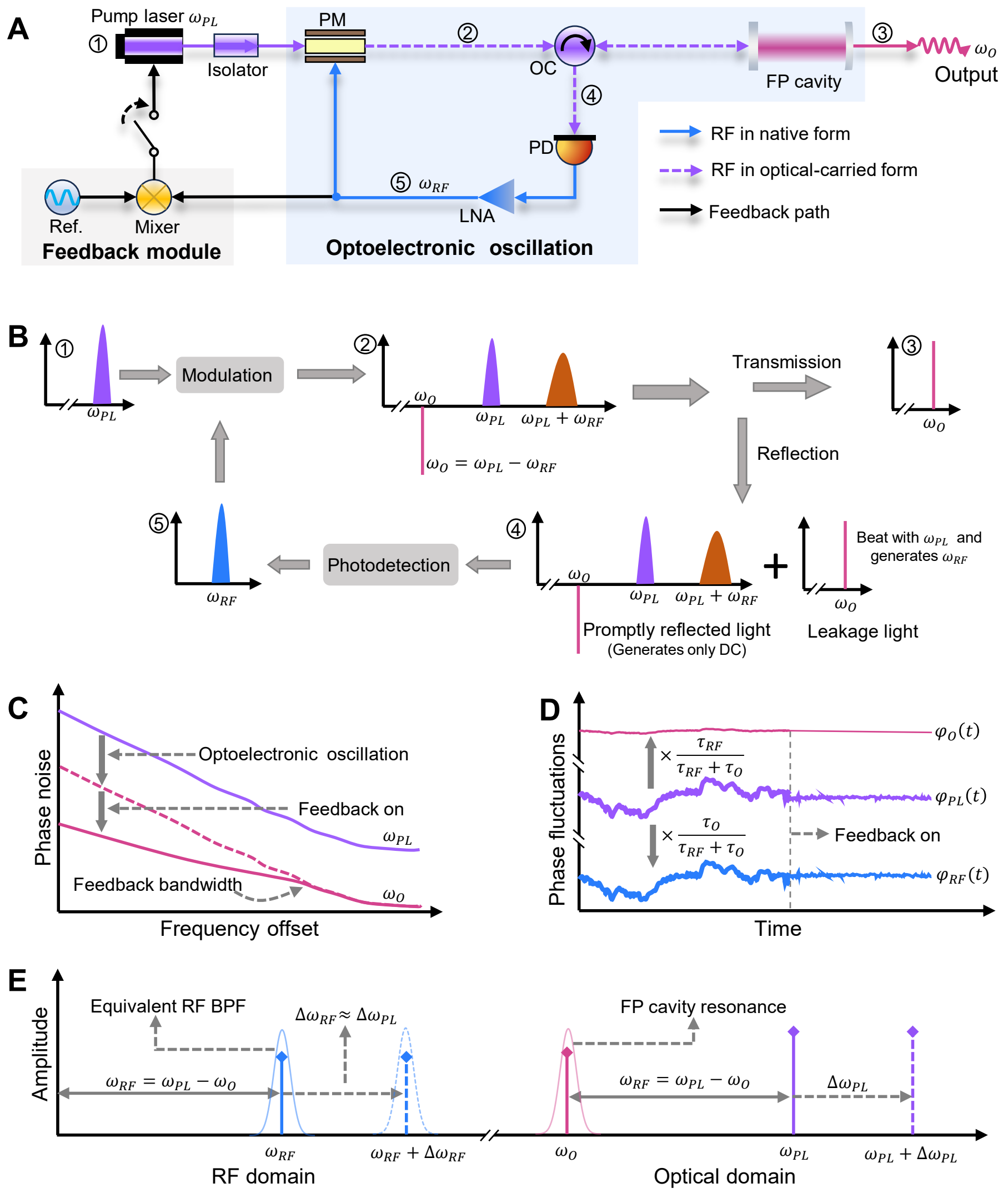

**Fig. 1. Schematic diagrams illustrating the concept of generating low-noise, narrow-linewidth light based on a PM-IM OEO system. (A)** A PM-IM OEO system with high-fineness FP cavity and shortened optoelectronic link. PM: phase modulator, PD: photodetector, LNA: low-noise amplifier, OC: optical circulator, Ref.: reference. **(B)** Spectra of the free-running PM-IM OEO. The spectrum width indicates the phase noise level. As the RF oscillation captures the phase noise of the PL, the phase modulation process generates a low-noise sideband and a noisy sideband with twice the phase noise of the PL. The opposite direction between the leaked and promptly reflected lower sidebands indicates a $\pi$-phase difference, which results in destructive interference in the total reflected field. **(C)** and **(D)** Illustration of dual-noise suppression mechanism in the frequency and time domains. **(E)** The equipment RF BPF based on PM-IM conversion using a narrow FP cavity. The frequency difference between the PL and the cavity resonance determines the gain window for the RF. This leads to the frequency drift of the generated RF following that of the PL, leading to one of the sidebands (the desired optical oscillation) roughly self-aligns with the cavity resonance.

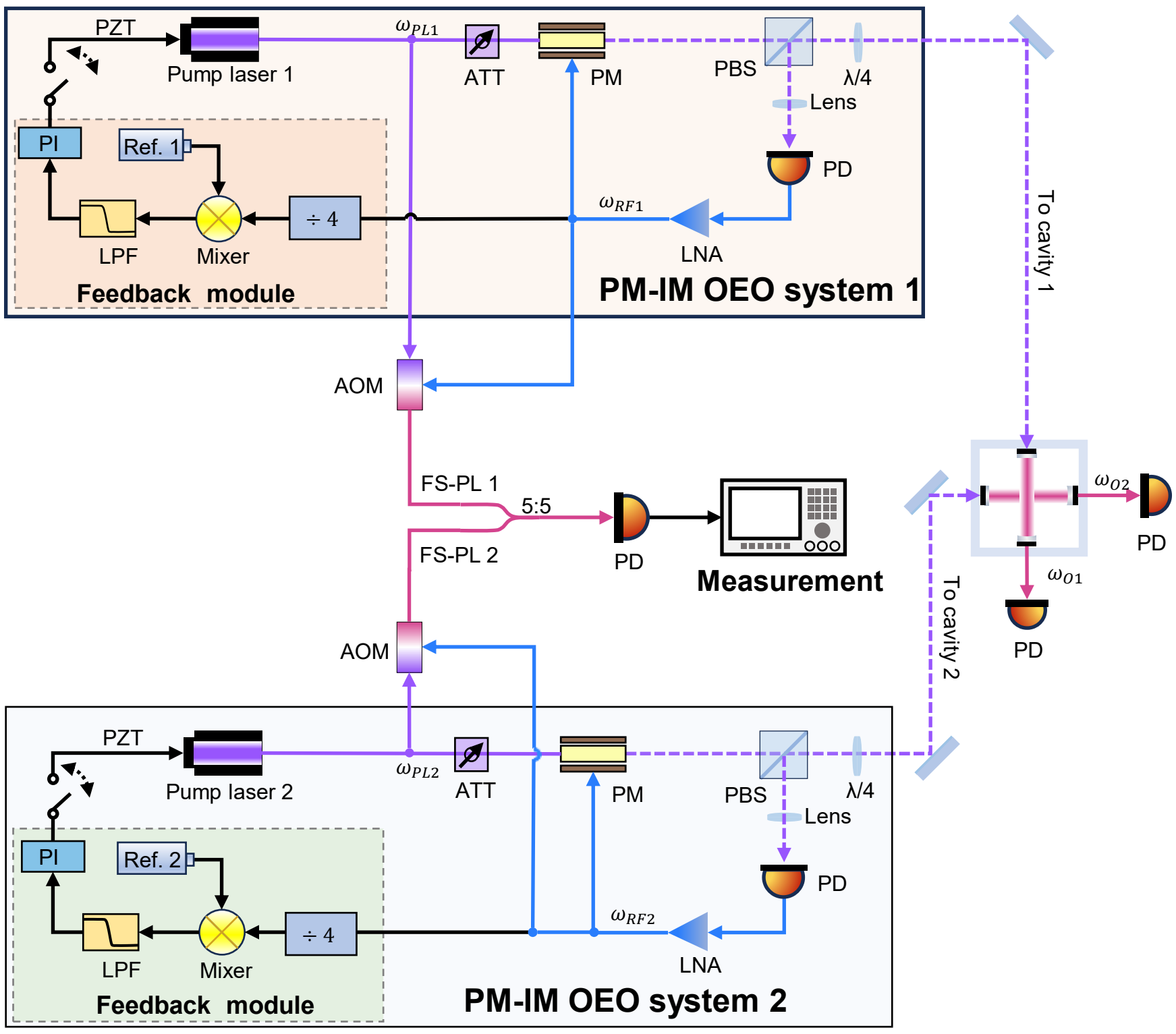

**Fig. 2. Experiment setup of the OEO-locking scheme.** Two PM-IM OEO systems based on two high-fineness FP cavities are built. The two FP cavities are housed within the same vacuum chamber. Due to the low coupling efficiency, the optical oscillation output power from the FP cavity's transmission port is small. As a workaround, the RFs are used to frequency-shift the PLs to obtain low-noise, narrow-linewidth lights. PBS: polarization beam splitter. PZT: piezoelectric ceramics, bandwidth of 200 kHz. LPF: low-pass filter, cut-off frequency 1 MHz. PI: Proportional-integral controller. AOM: Acousto-optic modulator, centers at 400 MHz with bandwidth about 40 MHz. ATT: tunable attenuator.

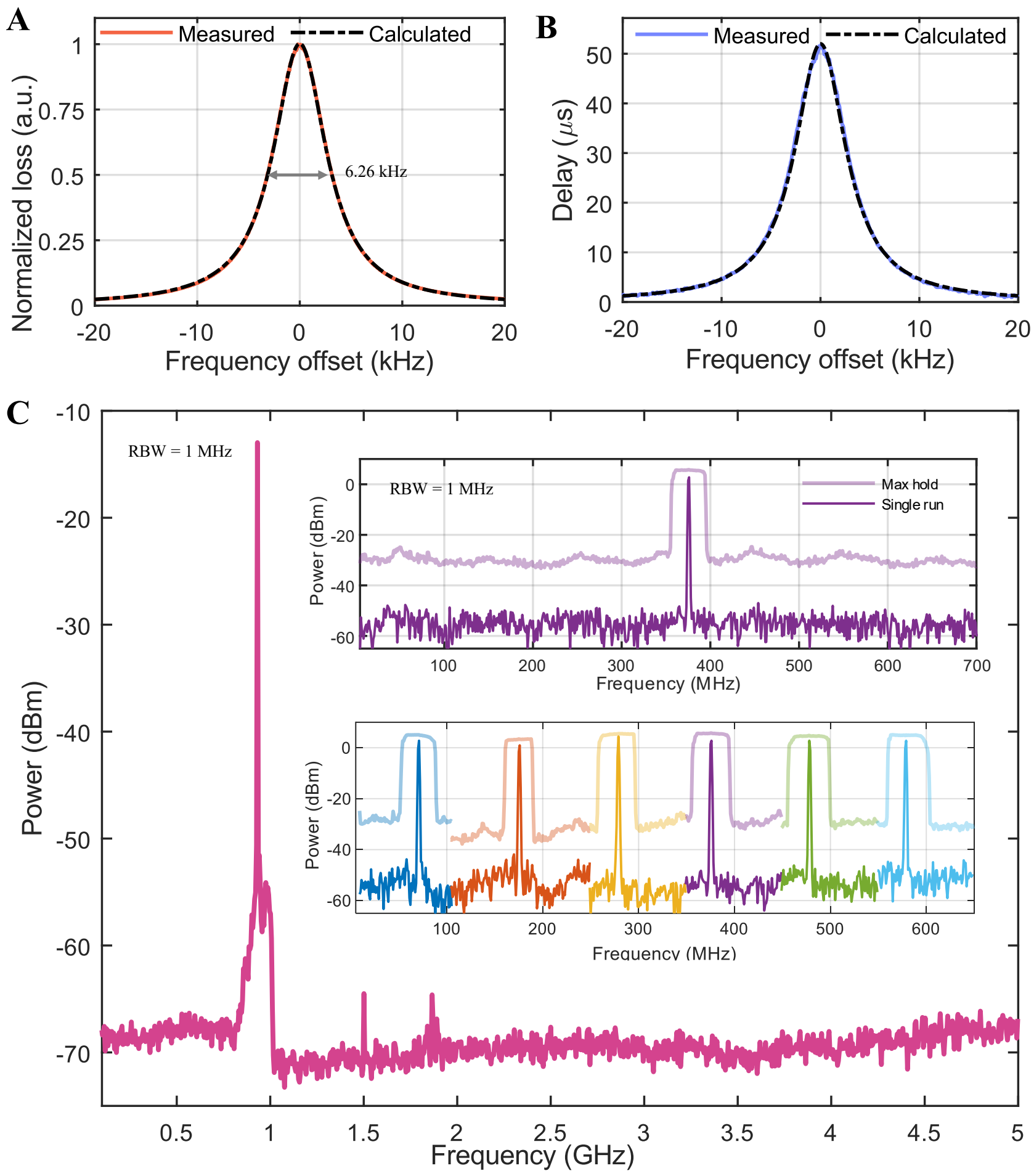

**Fig. 3. Experimental results.** The bandwidth **(A)** and group delay (**B**) of one equivalent RF BPF based on PM-IM conversion. The corresponding FP cavity has the same bandwidth and group delay as that of the RF BPF. The solid lines are experimentally measured results, and the dotted lines are theoretical calculations. **(C)** The power spectrum of the beat note between the two free-running FS-PL. Top inset: The power spectrum of one of the free-running RF signals. Bottom inset: OEO oscillates at different modes. The light-colored lines are measured in max-hold mode and the normal-colored lines are measured in single-run mode. RBW: resolution bandwidth.

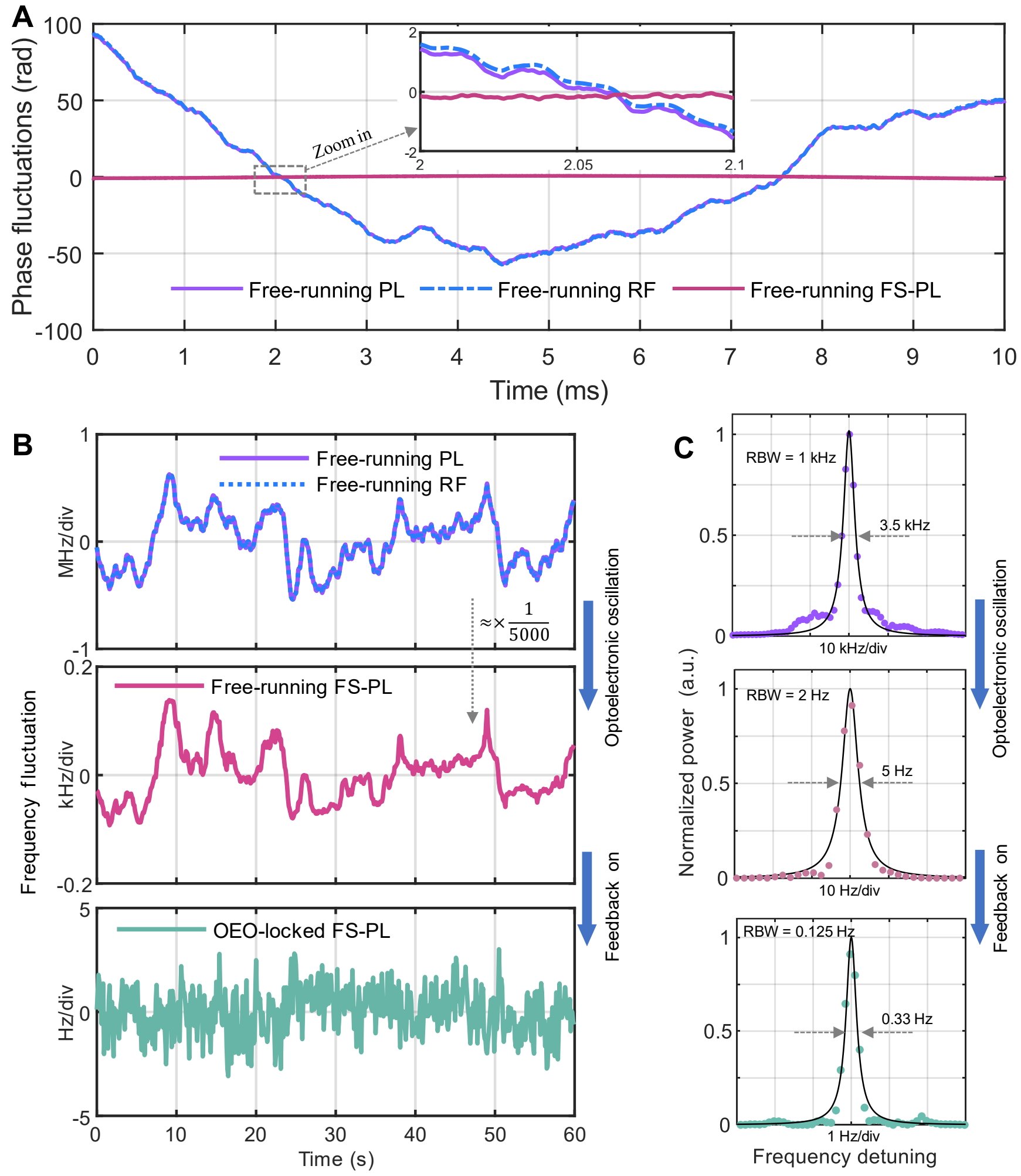

**Fig. 4. Experimental results. (A)** The phase fluctuations of the PL, RF signal, and the FS-PL in one of the free-running PM-IM OEOs. Inset: zoom-in view of the phase fluctuations. **(B)** The frequency fluctuations. Top: free-running PL and RF. Middle: the free-running FS-PL. Bottom: the beat signal between two OEO-locked FS-PLs. **(C)** The power spectra. Top: free-running PL. Middle: the free-running FS-PL. Bottom: the beat signal between two OEO-locked FS-PLs. The solid points are experimental data, while the black lines represent Lorentz line shape fitting.

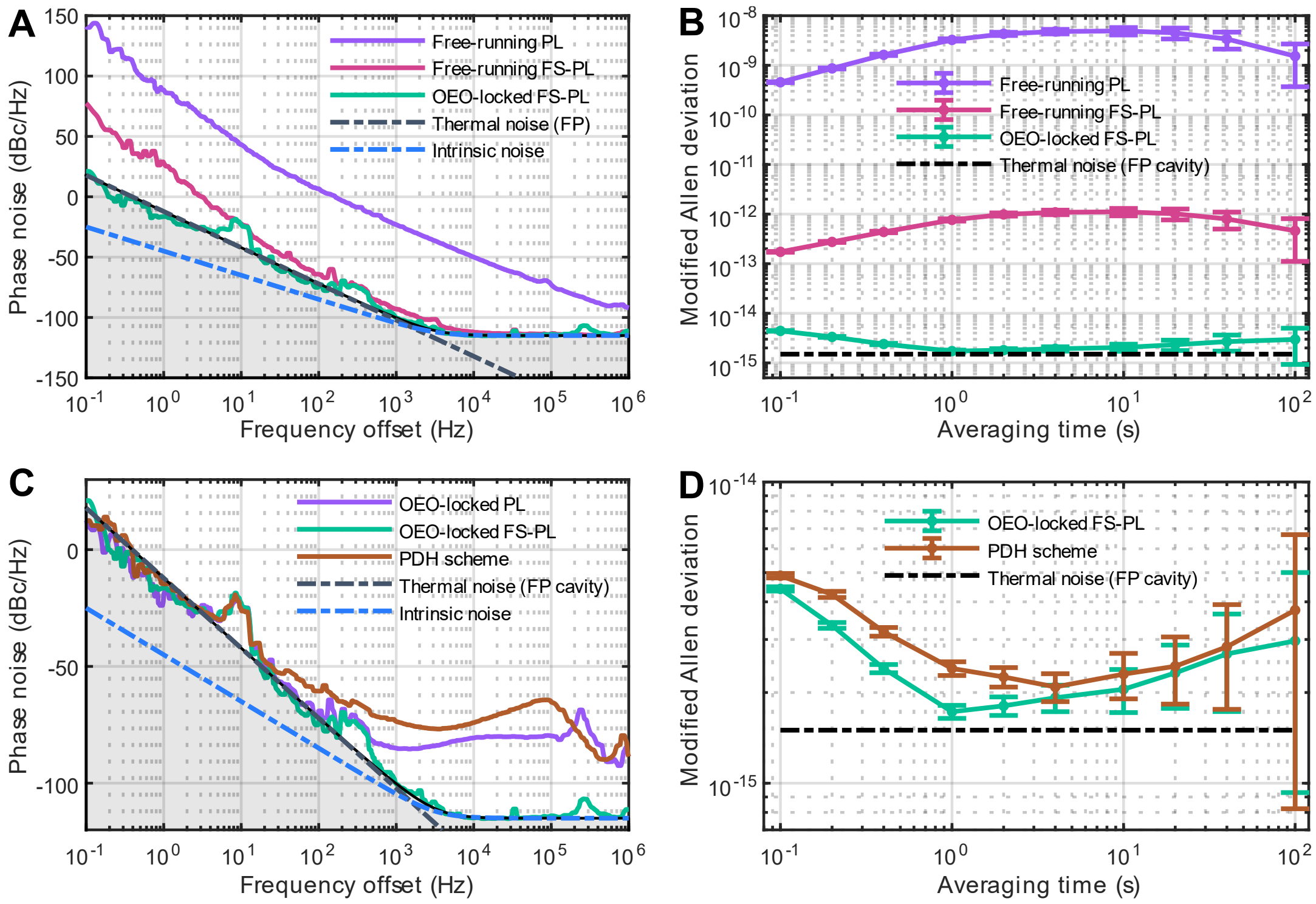

**Fig. 5 Experimental results. (A)** The comparison of the phase noise spectra between the free-running PL, the free-running FS-PL, and the OEO-locked FS-PL. The shadow indicates the limit set by the combination of the FP cavity's thermal noise and OEO's intrinsic noise. **(B)** The comparison of the modified Allan deviation between the free-running PL, the free-running FS-PL, and the OEO-locked FS-PL. **(C)** The comparison of the phase noise spectra between the OEO-locked PL, OEO-locked FS-PL, and the PDH scheme. **(D)** The comparison of the modified Allan deviation between the OEO-locked FS-PL and the PDH scheme. Linear frequency drifts have been subtracted from each data set.

Supplementary Information for

# Ultra-narrow linewidth light generation based on an optoelectronic oscillator

Qizhuang Cen[1,2,3,†], Shanhong Guan[1,2,3,†], Dongdong Jiao[4,5], Tengfei Hao[1,2,3], X. Steve Yao[6,7], Yitang Dai[8,*], Ming Li[1,2,3,*]

[1]Key Laboratory of Optoelectronic Materials and Devices, Institute of Semiconductors, Chinese Academy of Sciences; Beijing 100083, China.

[2]College of Materials Science and Opto-Electronic Technology, University of Chinese Academy of Sciences, Beijing 100049, China.

[3]School of Electronic, Electrical and Communication Engineering, University of Chinese Academy of Sciences, Beijing 100049, China

[4]National Time Service Center, Chinese Academy of Sciences; Xi'an 710600, China.

[5]Key Laboratory of Time and Frequency Standards, Chinese Academy of Sciences; Xi'an 710600, China.

[6]Photonics Information Innovation Center and Hebei Provincial Center for Optical Sensing Innovations, College of Physics Science & Technology, Hebei University; Baoding 071002, China.

[7]NuVision Photonics, Inc.; Las Vegas, NV 89109, USA.

[8]State Key Laboratory of Information Photonics and Optical Communications, Beijing University of Posts and Telecommunications; Beijing 100876, China.

†These authors contribute equally.

*Corresponding authors. Email: ytdai@bupt.edu.cn; ml@semi.ac.cn;

## I. Characterization of the FP cavities and the equivalent RF BPFs

### A. The transfer function of the leakage field

The field reflected from an Fabry–Pérot (FP) cavity is the coherent sum of the promptly reflected field from the first mirror and the leakage field, as shown in Fig. S1 (*23, 25*). The promptly reflected field is the immediate reflection from the first mirror, which does not enter the cavity. In contrast, the leakage field is a small portion of the field inside the cavity that escapes back through the first mirror. For simplicity, the FP cavity is modeled as lossless and symmetric, with the power transmission and reflection coefficients denoted as $T$ and $R$ ($T + R = 1$), respectively. Assuming $a_{In}$ is the incident field, the total reflected field can be expressed as:

$$\begin{aligned} a_{Out} &= -a_{In}r + a_{In}\left(Tre^{-i\phi} + Tre^{-i\phi}\cdot Re^{-i\phi} + \cdots Tre^{-i\phi}\cdot\left(Re^{-i\phi}\right)^{n} + \cdots\right) \\ &= -a_{In}r + a_{In}Tre^{i\phi}\sum_{1}^{n\to+\infty}\left(Re^{-i\phi}\right)^{n-1} \\ &= \underbrace{-a_{In}r}_{\text{Promptly reflected}} \underbrace{+a_{In}\frac{Tre^{-i\phi}}{1-Re^{-i\phi}}}_{\text{Leakage}} \\ &= a_{In}\frac{r\left(e^{-i\phi}-1\right)}{1-Re^{-i\phi}} \end{aligned} \tag{S1}$$

where $r$ is the amplitude reflection coefficient of the first mirror of the FP cavity, $\phi = \frac{2\pi f}{f_{FSR}}$ is the phase shift of one roundtrip, and $f_{FSR}$ is the free spectra range of the FP cavity. The minus sign in the promptly reflected field indicates a phase shift of $\pi$ as the optical field is reflected in going from vacuum to the mirror. Here, $\frac{r\left(e^{-i\phi}-1\right)}{1-Re^{-i\phi}}$ is well-known as the transfer function of the FP cavity for the total reflected field. The transfer function of the FP cavity for the leakage field is given by:

$$H_{FP}\left(f\right) = \frac{Tre^{-i\phi}}{1-Re^{-i\phi}} \tag{S2}$$

When the incident field is in resonance with the cavity, the leakage field and the promptly reflected field undergo a destructive interface, resulting in a notch in the total reflected spectrum. The transfer function for the leakage field shares the same normalized form as that for the transmission field, implying identical bandwidth and group delay characteristics. The 3-dB bandwidth of the $\left|H_{FP}\left(f\right)\right|^2$ is given as $\Delta f_{BW} = \frac{f_{FSR}}{F}$, where $F$ is the fineness of the FP cavity. The group delay of the FP cavity at the offset frequency $f$ from the resonance is given as:

$$\tau_{FP}\left(f\right) = -\frac{d\left\{\arg\left[H_{FP}\left(j2\pi f\right)\right]\right\}}{2\pi df} = \frac{1 - R\cos\left(\phi\right)}{1 + R^2 - 2R\cos\left(\phi\right)} \cdot \frac{1}{f_{FSR}}. \tag{S3}$$

At the cavity resonance, the group delay is $\tau_{FP}\left(0\right) = \frac{1}{Tf_{FSR}}$. Consequently, the delay-bandwidth product (DBWP) of the FP cavity for the leakage field can thus be calculated as $\tau_{FP}\left(0\right) \cdot \Delta f_{BW} = \frac{1}{\pi}$. When optical oscillation aligns with the cavity resonance, we obtain the group delay as $\tau_O = \tau_{FP}\left(0\right) = \frac{1}{\pi \Delta f_{BW}}$.

### B. Measurement of the equivalent RF BPF based on PM-IM conversion

The bandwidths and group delays of the FP cavities were obtained by measuring the equivalent radio frequency (RF) bandpass filters (BPFs) based on the phase-modulation-to-intensity-modulation (PM-IM) conversion using these cavities, as shown in Fig. S2. The equivalent RF filters, utilizing microwave photonic techniques, are also known as microwave photonic filters (MPF) (29). The MPF output is the beat signal between the promptly reflected light and the leakage light from the FP cavity. Due to the narrow bandwidth of the FP filter, the leakage light consists of either a single first-order sideband or none, contingent on the frequency difference between the laser and the FP cavity resonance. Mathematically, a laser with angular frequency $\omega_{PL}$, phase-modulated by an RF signal with angular frequency $\omega_{RF}$ under small modulation index $\beta$, can be expressed as $E_{PM}\left(t\right) = e^{-i\left[\omega_{PL}t + \beta\cos\left(\omega_{RF}t\right)\right]}$. This phase-modulated light hits on the FP cavity and is reflected. By normalizing the responsivity and impedance of the photodetector (PD), the recovery of the RF signal from the photodetection of the reflected light can be expressed as:

$$
\begin{aligned}
E_{PD}(t) &= \Big| \underbrace{-rE_{PM}(t)}_{\text{Promptly reflected}} + \underbrace{E_{PM}(t) \otimes h_{FP}(t)}_{\text{Leakage}} \Big|^2 \\
&\approx \left| -J_0(\beta) e^{-i\omega_{PL}t} + iJ_1(\beta) e^{-i(\omega_{PL}t - \omega_{RF})} \otimes h_{FP}(t) \right|^2 \\
&= -rJ_0 J_1 \operatorname{Re}\left\{ ie^{i(\omega_{PL}t - \omega_{RF})} \otimes h_{FP}(t) \cdot e^{-i\omega_{PL}t} \right\}
\end{aligned} \tag{S4}
$$

where $\otimes$ is the convolution operator and $h_{FP}(t)$ is the transfer function of the resonator for the leakage field. Equation S4 reveals how the equivalent RF BPF works: the RF signal is firstly up-converted by the pump laser (PL) and then passes through the optical filter with transfer function $h_{FP}(t)$, and finally down-converted by the same PL. If the PL is stable, the response of the RF BPF mirrors that of the FP cavity, albeit in the RF domain, as expressed by:

$$
S_{MPF}^{21}(f) = H_{FP}(f - f_{PL}) \tag{S5}
$$

Equation S5 suggests that properties of the FP cavity can be deduced by measuring the corresponding RF BPF.

In the experiment, we used an RF vector network analyzer (VNA) to measure the response of the RF BPFs, as shown in Fig. S2A. The setup involved an RF signal phase-modulating a continuous-wave stable laser, which then interacted with the FP cavity. As the modulation frequency is varied, one of the first-order sidebands swept across the FP cavity's resonance. This enabled the recovery of an RF signal capturing both amplitude and phase information of the FP cavity, providing valuable insights into its characteristics. By analyzing the transfer function of the PM-IM-based RF BPF, we were able to determine the characteristics of the FP cavities. Figure S3 displays the responses of two RF BPFs, exhibiting bandwidths of 6.26 kHz and 6.15 kHz, along with corresponding delays of 50.8 μs and 51.7 μs, respectively. These experimental results closely align with our theoretical predictions, confirming the FP cavities exhibit the same bandwidths and group delays.

## II. Input-output phase noise model of the free-running PM-IM OEO

In this section, we will explore the input-output phase noise model of the free-running PM-IM optoelectronic oscillator (OEO). The primary noise sources of the free-running PM-IM OEO come from the PL and the active optoelectronic link. Assuming that the frequency of the optical oscillation is lower than that of the PL, i.e., the lower sideband aligns with the resonance, the input-output phase-noise model of the OEO can described in Fig. S4 (32). Here, $\Phi_{PL}(s)$, $\Phi_{Intr}(s)$, $\Phi_O(s)$ and $\Phi_{RF}(s)$ are the Laplace transformation of the phase noise from the PL, the equipment input phase noise from optoelectronic link, the optical oscillation, and the RF oscillation. For stationary oscillation, these components are as follows:

$$\begin{cases} \Phi_{RF}(s) = \dfrac{1-H_{FP}(s)}{1-H_{FP}(s)e^{-s\tau_{RF}}}\Phi_{PL}(s) - \dfrac{1}{1-H_{FP}(s)e^{-s\tau_{RF}}}\Phi_{Intr}(s) \\ \Phi_O(s) = \dfrac{H_{FP}(s)\left(1-e^{-s\tau_{RF}}\right)}{1-H_{FP}(s)e^{-s\tau_{RF}}}\Phi_{PL}(s) - \dfrac{H_{FP}(s)e^{-s\tau_{RF}}}{1-H_{FP}(s)e^{-s\tau_{RF}}}\Phi_{Intr}(s) \end{cases}, \tag{S6}$$

where $s = i2\pi f$, and $f$ is offset frequency. The transfer functions of $H_{PL-RF}(s) = \dfrac{1-H_{FP}(s)}{1-H_{FP}(s)e^{-s\tau_{RF}}}$ and $H_{PL-O}(s) = \dfrac{H_{FP}(s)\left(1-e^{-s\tau_{RF}}\right)}{1-H_{FP}(s)e^{-s\tau_{RF}}}$ describe how phase fluctuations from the PL are transferred to the RF and optical oscillations, respectively. This relationship outlines a framework for understanding the dynamic interactions between these oscillations within the PM-IM OEO system. The second terms in $\Phi_O(s)$ and $\Phi_{RF}(s)$ describe fundamental phase noise set by the intrinsic noise in the active optoelectronic link.

### A. Distribution of the PL phase fluctuations

When considering a frequency offset range that is significantly smaller than the free spectra range (FSR) of the FP cavity, i.e., $f < f_{FSR}$, the amplitude and phase response of the transfer function $H_{PL-RF}(s)$ can be approximated as constant with $\left|H_{PL-RF}(s)\right| = \dfrac{\tau_O}{\tau_O + \tau_{RF}}$ and $Ang\left[H_{PL-RF}(s)\right] = 0$. This implies that the PL's phase fluctuations are linearly transferred to the RF with a constant coefficient $\dfrac{\tau_O}{\tau_O + \tau_{RF}}$.

Given that $\tau_O \gg \tau_{RF}$, the transfer coefficient is very close to unity, i.e., $\frac{\tau_O}{\tau_O+\tau_{RF}} \approx 1$, suggesting that the PL's phase fluctuations are accurately replicated in the RF domain. This efficient, broadband, and linear extraction of phase fluctuations facilitates prompt and precise feedback. With the experimental parameters of $\tau_O = 50.8$ μs and $\tau_{RF} = 10$ ns, we calculated the transfer function $H_{PL-RF}(s)$, and the results shown in Fig. S5 confirm the accuracy of these approximations in both amplitude and phase.

Similarly, within the bandwidth of the FP cavity, the amplitude and phase responses of the transfer function $H_{PL-O}(s)$ can be approximated as constant with $\left|H_{PL-O}(s)\right| = \frac{\tau_{RF}}{\tau_O+\tau_{RF}} \approx 0$ and $Ang\left[H_{PL-O}(s)\right] = 0$, respectively. Calculations performed under the same experimental parameters validate these approximations with high accuracy, as displayed in Fig. S6. As a result, the transfer coefficients of phase fluctuations from the PL to the RF and the optical oscillation in the time domain can be approximately expressed by:

$$\begin{cases} \varphi_{RF}(t) = \dfrac{\tau_O}{\tau_{RF}+\tau_O}\varphi_{PL}(t) \\ \varphi_O(t) = \dfrac{\tau_{RF}}{\tau_{RF}+\tau_O}\varphi_{PL}(t) \end{cases} \tag{S7}$$

where $\varphi_{PL}(t)$, $\varphi_{RF}(t)$, and $\varphi_O(t)$ are the time-domain phase fluctuations of the PL, the RF oscillation, and the optical oscillation, respectively. In terms of the power spectral density (PSD) of the phase noise, the transfer coefficients in dB are $20\log_{10}\left(\frac{\tau_O}{\tau_O+\tau_{RF}}\right)$ and $20\log_{10}\left(\frac{\tau_{RF}}{\tau_O+\tau_{RF}}\right)$, respectively.

### B. Intrinsic noise in the PM-IM OEO

The intrinsic noise in the OEO primarily arises from the active optoelectronic link, including thermal noise, PL relative intensity noise (RIN), and shot noise (*28*). The total noise density can be quantified as follows:

$$\rho_N = 4k_B T_0 (NF) + \langle I_{PD}\rangle^2 N_{RIN} Z + 2e\langle I_{PD}\rangle Z, \tag{S8}$$

where $k_B = 1.38\times10^{-23}$ *J/K* is Boltzmann's constant, $T_0 = 290$ *K* is the ambient temperature, *NF* is the noise factor of the RF amplifier, $\langle I_{PD}\rangle$ is the average photocurrent, $N_{RIN}$ is the laser RIN, $e = 1.6\times10^{-19}c$ is the charge of an electron, and *Z* is the impedance. Assuming the intrinsic noise is white noise, it impacts both phase noise and intensity noise equally. The equipment input phase noise of the intrinsic noise can be expressed as $\left|\Phi_{Intr}(s)\right|^2 = \frac{G_{Amp}^2 \rho_N}{2P_{RF}}$, where $G_{Amp}$ is the voltage gain of the RF amplifier, $P_{RF}$ is the RF power after the LNA. The phase noise of the generated RF contributed by the intrinsic noise can be given by

$$S_{Intr\text{-}RF}(f) = \left|\frac{1}{1-H_{FP}(s)e^{-s\tau_{RF}}}\Phi_{Intr}(s)\right|^2 \approx \frac{G_{Amp}^2\rho_N}{2P_{RF}}\left(\frac{1}{(2\pi f\tau_{RT})^2}+1\right), \quad \text{(S9)}$$

where $\tau_{RT} = \tau_{RF} + \tau_O$ is considered as the equivalent roundtrip time of the optoelectronic cavity. The phase noise of the optical oscillation contributed by the intrinsic noise can be expressed as:

$$S_{Intr\text{-}O}(f) = \left|\frac{H_{FP}(s)e^{-s\tau_{RF}}}{1-H_{FP}(s)e^{-s\tau_{RF}}}\Phi_{Intr}(s)\right|^2 \approx \frac{G_{Amp}^2\rho_N}{2P_{RF}(2\pi f\tau_{RT})^2}. \quad \text{(S10)}$$

The approximation in Eq. S10 is based on $\tau_O \approx \tau_{RT}$. And $\frac{G_{Amp}^2\rho_N}{P_{RF}}$ is considered as the noise-to-signal ratio and is roughly estimated by measuring the power spectrum of the RF signal. The noise-to-signal ratio is measured to be −112 dBc/Hz. Given a roundtrip time of $\tau_{RT} = 50.8\ \mu s$, the phase noise contributed by the intrinsic noise for both the generated RF and optical oscillation is calculated and presented in Fig. S7.

The phase noise of the RF contributed from the intrinsic noise shows a dependency of $-20log_{10}(f)$ at low-frequency offsets, a characteristic typical of noise accumulation in a feedback oscillator. However, due to cavity filtering, the intrinsic-noise-induced phase noise in the optical oscillation maintains this dependency across the entire observed frequency range. At low-frequency offsets, the phase noise contributed by intrinsic noise is negligible compared to the thermal noise of the FP cavity, but it becomes dominant at high-frequency offsets. The integrated linewidth of the intrinsic noise is calculated to be less than 40 μHz, suggesting that the linewidth is limited by the thermal noise of the FP cavity.

In the current experimental setup, directly collecting optical oscillation from the transmission of the FP cavity is challenging due to the low coupling coefficient.

Therefore, we obtain low-noise light by frequency-shifting the PL with the generated RF. The FS-PL exhibits a flat floor at higher frequency offsets, as shown in Fig. 5A and Fig. S7. The setup, hindered by low coupling efficiency between the optoelectronic link and the FP cavity, necessitates a large RF power gain, which increases the noise-to-signal ratio. By utilizing a low-$V_\pi$ modulator and minimizing losses in the optoelectronic link, we can reduce the electric gain, thereby decreasing the noise-to-signal ratio and reducing phase noise introduced by the intrinsic noise. With further optimization of optoelectronic link parameters, the phase noise induced by the intrinsic noise can reach a minimal value that is equal to half of the laser RIN (*28*).

## III. Suppression limit of PL noise in free-running OEO

To significantly enhance phase noise suppression, one effective strategy is to use a narrow FP cavity by increasing the finesse and/or reducing its FSR. Additionally, minimizing the delay in the optoelectronic link also enhances the suppression capacity achievable using high-speed, integrated optoelectronic devices combined with advanced microassembly techniques. However, to achieve single-frequency oscillation in the RF and optical domains, it is essential to incorporate an electrical low-pass filter (LPF) with a cutoff frequency that is less than half the FSR of the FP cavity. The cumulative delay, which includes contributions from various optical and electrical components, can be expressed as:

$$\tau_{RF} = \tau_{LPF} + \tau_{Amp} + \tau_{PM} + \tau_{PD} + \tau_{con}, \tag{S11}$$

where $\tau_{LPF}$, $\tau_{Amp}$, $\tau_{PM}$, $\tau_{PD}$, and $\tau_{con}$ are the delays or the response times of the LPF, electronic amplifier, phase modulator (PM), PD, and the connection between these components, respectively. Typically, the response time of high-speed PMs, PDs, and electronic amplifiers is generally a few tens of picoseconds. With integration and refined microassembly techniques, the connection length between these components can be reduced to a few centimeters, or even less, resulting in a connection delay of under 0.1 ns. Consequently, in a PM-IM OEO where the cavity FSR is much smaller than the bandwidths of the PD and the electronic amplifier, the LPF dominates the delay in the optoelectronic link. The transfer function of a first-order LPF can be described as $H_{LPF}(s) = \frac{2\pi f_c}{2\pi f_c + s}$, where $f_c$ is the cut-off frequency of the LPF and $f_c = \frac{1}{2} f_{FSR}$ to guarantee single-mode operation. The maximum delay of the LPF is $\tau_{\max-LPF} = \frac{1}{2\pi f_c} = \frac{1}{\pi f_{FSR}}$. We use this value to roughly estimate the suppression without feedback control. The suppression can be as large as $20\log_{10}(F)$ dB in PSD.

Assuming that the delays in the optoelectronic link are as follows: $\tau_{LPF} = \frac{1}{\pi \cdot f_{FSR}}$, $\tau_{Amp} = \tau_{PM} = \tau_{PD} = 20$ ps, and $\tau_{con} = 0.1$ ns, we compute the phase noise suppression as a function of the cavity finesse and FSR. Results shown in Fig. S8 indicate that the use of an FP cavity with a large FSR or relatively low finesse results in limited suppression due to the optoelectronic components, and the connections between these components mainly contributing to the delay. This typically occurs in PM-IM OEO systems that utilize integrated optical resonators. Conversely, using an FP cavity with a small FSR and high finesse leads to significant suppression. In this case, the LPF primarily contributes to the delay in the optoelectronic link. With parameters such as a 1-GHz FSR and finesse of $10^6$, the suppression of phase noise can reach up to 120 dB, corresponding to a frequency fluctuation suppression of $10^6$. In our experiments, the delay in the optoelectronic link, predominantly arising from the connections between discrete optoelectronic components, results in a limited frequency fluctuation suppression of 5000.

## IV. Frequency instability introduced by the optoelectronic cavity

In this section, we discuss the impact of delay fluctuations in the active optoelectronic link and the passive FP cavity on the frequency stability of oscillations. In feedback oscillators, fluctuations in the cavity delay affect the output frequency stability, which can be approximated by the formula $\Delta\omega_s = -\frac{\Delta\tau_{RT}}{\tau_{RT}}\omega_s$. Here, $\Delta\tau_{RT}$ represents the fluctuations in roundtrip time, while $\omega_s$ denotes the oscillation frequency. The delay fluctuations pose a significant challenge in directly generating high-frequency, low-noise signals, primarily due to the inherent delay instability in an active laser cavity. However, the relative instability, quantified as $\frac{\Delta\tau_{RT}}{\tau_{RT}}$, can be significantly reduced in a passive optical cavity. This stability enhancement facilitates the generation of ultra-narrow linewidth light by locking a laser to a passive optical resonator. The OEO operates as a feedback oscillator where delay fluctuations—whether in the active optoelectronic link or the passive FP cavity—induce frequency fluctuations in both the RF and the optical oscillations. For stable oscillations, the OEO must satisfy the phase condition of the Barkhausen stability criterion and maintain a specific frequency relationship between the PL and the oscillations. This relationship is expressed as:

$$\begin{cases} \phi_O \pm \phi_{RF} = 2N\pi \\ \omega_O \pm \omega_{RF} = \omega_{PL} \end{cases}, \tag{S12}$$

where $\phi_O$ and $\phi_{RF} = \omega_{RF}\tau_{RF}$ are the phase shifts introduced by the FP cavity and the optoelectronic link, $\omega_{PL}$, $\omega_O$ and $\omega_{RF}$ are the angular frequencies of the PL, the optical oscillation, and the RF, respectively. The signs "$\pm$" represent whether the lower or the upper sideband aligns to the FP cavity resonance, i.e., $\omega_O < \omega_{PL}$ and $\omega_O > \omega_{PL}$, respectively.

### A. Frequency instability in free-running OEO

In the free-running OEO, for simplicity, we assume the PL is stable and deduce how the optoelectronic cavity instability is transferred to RF and optical oscillations. Near the FP cavity resonance, the optical phase shift can be approximated as $\phi_O = \omega_O\tau_O$. When the delay, $\tau_{RF}$, is changed to $\tau_{RF} + \Delta\tau_{RF}$, the oscillating frequencies of both the optical and the RF domain are adjusted to meet the phase condition, as expressed by:

$$\begin{cases}(\omega_O+\Delta\omega_O)\tau_O \pm (\omega_{RF}+\Delta\omega_{RF})(\tau_{RF}+\Delta\tau_{RF}) = 2N\pi \\ \Delta\omega_O \pm \Delta\omega_{RF} = 0\end{cases}, \tag{S13}$$

where $\Delta\omega_O$ and $\Delta\omega_{RF}$ are the angular frequency fluctuations of the optical and RF oscillations, respectively. Ignoring higher order terms in approximations, the frequency instabilities of these two oscillations induced by $\Delta\tau_{RF}$ can be expressed as:

$$\begin{cases}\dfrac{\Delta\omega_O}{\omega_O} = \pm\dfrac{\Delta\tau_{RF}}{\tau_O+\tau_{RF}}\dfrac{\omega_{RF}}{\omega_O} \\ \dfrac{\Delta\omega_{RF}}{\omega_{RF}} = -\dfrac{\Delta\tau_{RF}}{\tau_O+\tau_{RF}}\end{cases}. \tag{S14}$$

Given that $\tau_O \gg \tau_{RF}$ and $\omega_O \gg \omega_{RF}$, the delay fluctuations $\Delta\tau_{RF}$ has a negligible impact on the frequency stability of optical oscillation. By utilizing a narrow FP cavity and minimizing $\tau_{RF}$—for instance, with a 6.4-kHz FP cavity ($\tau_O = 50\ \mu s$), an RF delay of $\tau_{RF} = 5\ ns$, and operating frequencies of 200 MHz for RF and 200 THz for the optical oscillation—the frequency instability $\frac{\Delta\omega_O}{\omega_O}$ can be reduced to a remarkable level of $10^{-19}$, assuming the instability of the active optoelectronic link is $10^{-9}$. Consequently, the frequency stability of optical oscillation is ultimately determined by that of the passive optical resonator.

Similarly, the frequency instantly caused by the passive FP cavity can be approximately expressed as:

$$\begin{cases}\dfrac{\Delta\omega_O}{\omega_O} = -\dfrac{\Delta\tau_O}{\tau_O+\tau_{RF}} \approx -\dfrac{\Delta L}{L} \\ \dfrac{\Delta\omega_{RF}}{\omega_{RF}} = \pm\dfrac{\Delta\tau_O}{\tau_O+\tau_{RF}}\dfrac{\omega_O}{\omega_{RF}} \approx \pm\dfrac{\Delta L}{L}\dfrac{\omega_O}{\omega_{RF}}\end{cases}, \tag{S15}$$

where $\Delta L$ is the length variation of the FP cavity. We can conclude that the frequency stability of optical oscillation closely follows that of the stability of the FP cavity length. The transfer coefficients of the delay instability to the frequency instability, as a function of the delay ratio, are calculated and shown in Fig. S9. The calculations suggest that in the OEO with a large ratio of $\tau_O{:}\,\tau_{RF}$, the optical oscillation closely follows the stability of the FP cavity and is nearly immune to the instability of the active optoelectronic link. Conversely, the instability of the FP cavity is amplified in the RF domain.

### B. Frequency instability in OEO with feedback

When the feedback is on, the OEO-generated RF is locked to the stable external reference, resulting in $\Delta\omega_{RF}=0$. The delay fluctuations $\Delta\tau_{RF}$ in the optoelectronic cavity result in a frequency shift of the optical oscillation, which is satisfied with $(\omega_O+\Delta\omega_O)\tau_O \pm \omega_{RF}(\tau_{RF}+\Delta\tau_{RF})=2N\pi$. Based on this, the frequency instability caused by the $\Delta\tau_{RF}$ is expressed as $\frac{\Delta\omega_O}{\omega_O}=\mp\frac{\Delta\tau_{RF}}{\tau_O}\frac{\omega_{RF}}{\omega_O}$. Similarly, the frequency instability caused by the $\Delta\tau_O$ is expressed as $\frac{\Delta\omega_O}{\omega_O}=-\frac{\Delta\tau_O}{\tau_O}$. Here, we conclude that the cavity-induced frequency instability of optical oscillation behaves consistently in both the free-running and locked OEO. Specifically, the frequency instability predominantly depends on the passive FP cavity, while the impact of the active optoelectronic link on this instability is negligible.

### C. Tolerance for frequency drift of pump laser

Frequency drift in the PL leads to a corresponding drift in the optical oscillation frequency. This drift reduces the optical power of the leakage light, decreasing RF gain and potentially ceasing oscillation, a phenomenon known as quenching. Assuming a net power gain at the resonance of 3 dB, oscillation will quench when the frequency drift exceeds the bandwidth of the FP cavity. This implies that the frequency span of optical oscillation is equal to the bandwidth of the FP cavity. The corresponding phase shift of the optical oscillation, related to this frequency drifts $f_{BW}$, is $\pi/2$. This phase shift in optical oscillation is equal to that of the RF oscillation, expressed as $2\pi\Delta f_{RF}\tau_{RF}$. As the absolute frequency drift of RF oscillation is almost the same as that of the PL, the tolerance span of the PL drift is about $\Delta f_{Tol}\approx\Delta f_{RF}=\frac{1}{4\tau_{RF}}$. This suggests that a shorter delay in optoelectronic link results in a wider frequency span for stable oscillation. The use of high-speed, integrated devices and microassembly techniques can decrease the $\tau_{RF}$ down to 1 ns, thus supporting the tolerance of frequency drift over hundred MHz. Such a wide frequency span indicates that the OEO system is robust against the frequency drift of the PL, ensuring stable, continuous, low-noise light generation. This robustness also simplifies the implementation of feedback mechanisms, facilitating the development of a stable laser system.

## V. A comparison between the OEO-locking scheme and the PDH scheme

In this section, we provide a comparative analysis of noise performance between the OEO-locking scheme and the Pound–Drever–Hall (PDH)scheme. This analysis focuses on the output noise characteristics of the PDH system compared to those of the OEO-locking system. The objective is to demonstrate that the OEO-locking scheme can achieve superior noise suppression compared to the PDH scheme through its feedback mechanism and that it further provides broadband suppression with a large suppression factor. The phase noise at the output of from both systems can be categorized into three sources: residual PL noise, intrinsic noise from the error extraction link, and noise from the feedback link. These noise sources may prevent the output from reaching the thermal noise limit of the FP cavity. It is critical to note that while both systems are affected by similar types of noise, the level of these noises may be different due to factors such as optical power at the photodetector and link gain. In this analysis, we will focus on how each system suppresses or amplifies these noise sources, rather than quantifying the specific values of the noise level.

Among these noise sources, the phase noise of the PL is particularly critical. If the lasers in both schemes have identical phase noise levels, the residual phase noise at the output primarily depends on each system's ability to suppress noise. The intrinsic noise in both systems arises from the active optoelectronic link used for frequency or phase error extraction and includes thermal noise, laser RIN, and shot noise. Noise from the feedback link also undergoes similar processes, primarily influenced by the proportional and integral circuits used. We note here that an additional proportional-integrated (PI) circuit is required in the feedback link of the PDH system to achieve an equivalent noise suppression factor for the PL, considering that the error signal within the cavity bandwidth in the PDH system represents a frequency error, while it is a phase error in the OEO-locking system. For simplicity, we will treat the noise sources in the feedback link as cumulative noise, rather than analyzing individual contributions. This approach simplifies comparative analysis while focusing on the performance differences between the two systems.

### A. Input-output phase-noise model

The input-output phase noise of the PDH scheme and the locked OEO can be modeled in Fig. S10 (*32*). The detailed parameters and the transfer functions of the different components or subsystems are listed in Tables S1 and S2. A match delay equal to $\tau_{RF}$ is introduced to reach a perfect frequency shifting of the PL. The device parameters and the transfer functions are the same in these systems. Here, we define the transfer functions of the feedback links in the PDH and OEO-locking systems as

$H_{FB}^{PDH}(s)=H_{LPF}(s)H_{PI}^{PDH}(s)H_{PL}(s)$ and $H_{FB}^{OEO}(s)=H_{LPF}(s)H_{PI}^{OEO}(s)H_{PL}(s)$ , respectively. At the steady state, the phase noises of the output can be expressed as:

$$\begin{cases} \Phi_{PDH}(s)=\dfrac{1}{1-H_{FB}^{PDH}(s)D_{RF}(s)\left[1-H_{FP}(s)\right]}\Phi_{PL}(s) \\ \qquad +\dfrac{H_{FB}^{PDH}(s)D_{RF}(s)}{1-H_{FB}^{PDH}(s)D_{RF}(s)\left[1-H_{FP}(s)\right]}\Phi_{Intr}(s) \\ \qquad +\dfrac{H_{FB}^{PDH}(s)}{1-H_{FB}^{PDH}(s)D_{RF}(s)\left[1-H_{FP}(s)\right]}\Phi_{FB}(s) \\ \Phi_{PL}^{L}=\dfrac{1-H_{FP}(s)D_{RF}(s)}{1-\left[H_{FP}(s)+H_{FB}^{OEO}(s)H_{FP}(s)-H_{FB}^{OEO}(s)\right]D_{RF}}\Phi_{PL}(s) \\ \qquad -\dfrac{H_{FB}(s)D_{RF}(s)}{1-\left[H_{FP}(s)+H_{FB}^{OEO}(s)H_{FP}(s)-H_{FB}^{OEO}(s)\right]D_{RF}}\Phi_{Intr}(s) \\ \qquad +\dfrac{H_{FB}^{OEO}(s)\left[1-H_{FP}(s)D_{RF}(s)\right]}{1-\left[H_{FP}(s)+H_{FB}^{OEO}(s)H_{FP}(s)-H_{FB}^{OEO}(s)\right]D_{RF}}\Phi_{FB}(s) \\ \Phi_{FS-PL}^{L}=\dfrac{\left[1-D_{RF}(s)\right]H_{FP}(s)D_{RF}(s)}{1-\left[H_{FP}(s)+H_{FB}^{OEO}(s)H_{FP}(s)-H_{FB}^{OEO}(s)\right]D_{RF}}\Phi_{PL}(s) \\ \qquad -\dfrac{\left[1+H_{FB}^{OEO}(s)D_{RF}(s)\right]D_{RF}(s)}{1-\left[H_{FP}(s)+H_{FB}^{OEO}(s)H_{FP}(s)-H_{FB}^{OEO}(s)\right]D_{RF}}\Phi_{Intr}(s) \\ \qquad +\dfrac{\left(1-D_{RF}\right)H_{FB}(s)H_{FP}(s)D_{RF}(s)}{1-\left[H_{FP}(s)+H_{FB}^{OEO}(s)H_{FP}(s)-H_{FB}^{OEO}(s)\right]D_{RF}}\Phi_{FB}(s) \end{cases} \tag{S16}$$

It is important to highlight that the actual values of these noise components may vary between the two systems. This analysis focuses on comparing how each system processes these noise sources using their respective feedback mechanisms.

## B. Simulations

Based on Eq. S16 and specific parameters listed in Table S2, we calculated the phase noise suppression of the outputs in these systems, with the results presented in Fig. S11. Thanks to the additional integration process, the PDH scheme achieved the same suppression of the laser noise as that of the OEO-locked PL. However, the OEO-locked FS-PL benefits from an additional suppression factor of $20log_{10}\frac{\tau_{RF}}{\tau_{RT}}$ owing to the oscillation mechanism. This suppression spans a broad frequency range, constrained only by the bandwidth of the optoelectronic link as discussed in Section II. This dual-suppression mechanism integrating the feedback loop and the oscillation provides large and broadband noise suppression, outperforming other schemes. This dual-suppression mechanism also decreases the dependency on the PL's noise performance and the fineness of the optical cavity.

The impact of the intrinsic noise in the error extraction process is nearly identical between the PDH system and the OEO-locked PL, as illustrated in Fig. S11 B. However, at high-frequency offsets, the OEO-locked FS-PL exhibits a flat but higher noise level. We point out that the descent of high-frequency offsets in the PDH system and the OEO-locked PL comes from the low-pass filtering of the feedback link. In other systems where optical oscillation can be directly obtained from the transmission port of the optical resonator with adequate power, high-frequency noise is naturally filtered by the optical resonator itself.

The noise in feedback link impacts the PDH system and the OEO-locking system differently. In the PDH system, feedback link noise at low-frequency offsets presents primarily as frequency noise, similar to the intrinsic noise in the error extraction link. Conversely, in the OEO-locking system, feedback link noise across the observed frequency span manifests as phase noise, contributing uniformly to the phase noise profile. This analysis indicates that the PDH scheme is more vulnerable to feedback noise than the OEO-locking scheme. Furthermore, in the OEO-locked FS-PL , the oscillation mechanism effectively mitigates feedback noise, offering a distinct advantage in noise management compared to the PDH system. With significant suppression, feedback noise becomes negligible, allowing the optical oscillation to approach the limits set by the thermal noise of the FP cavity and the intrinsic noise of the optoelectronic link.

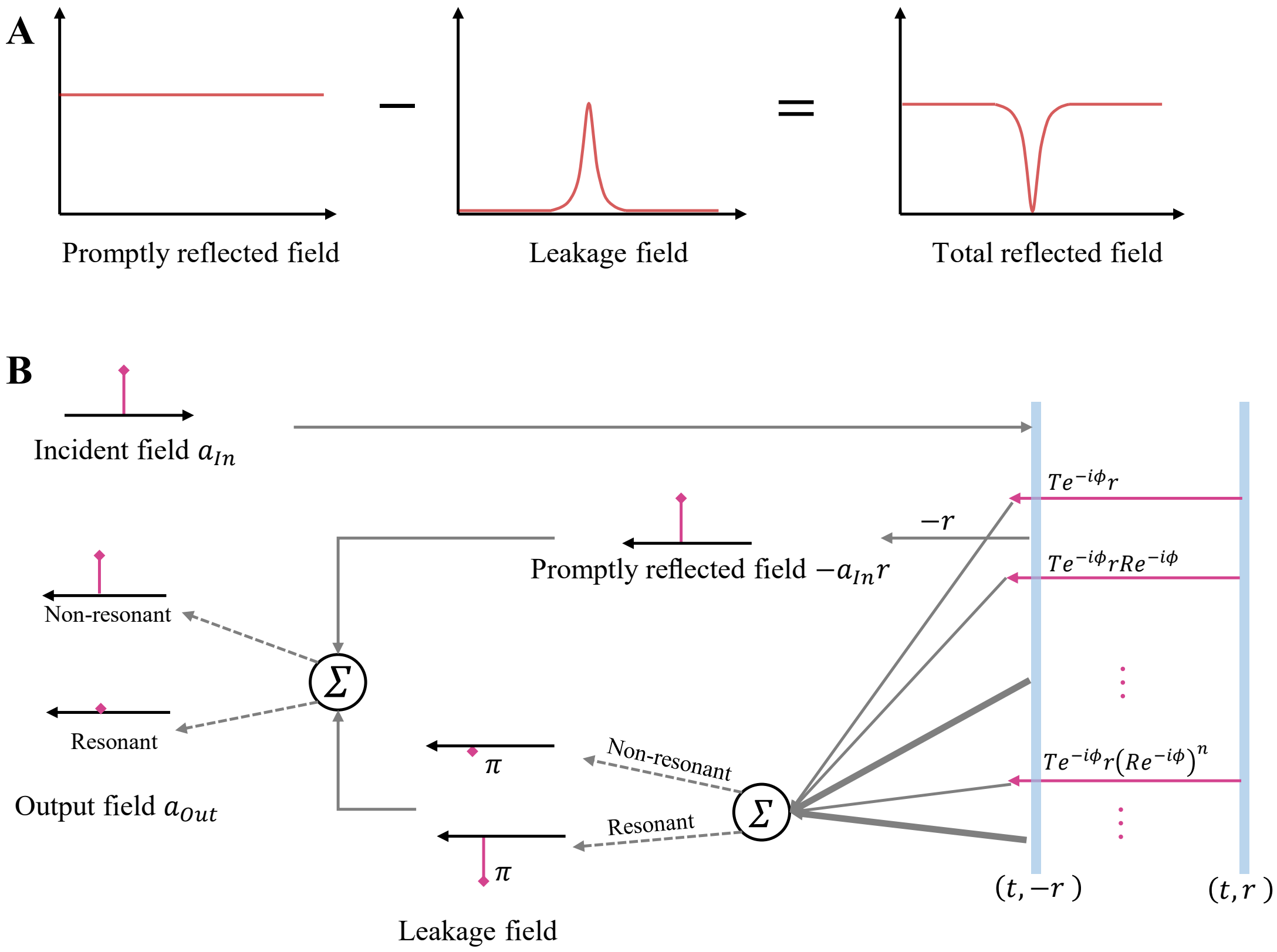

**Fig. S1. The schematic diagram transfer function of the total reflected field of an FP cavity. (A)** The notch is the result of the destructive interference between the promptly reflected field and the leakage field. Based on Stokes relationship, the promptly reflected field and the leakage field have a phase difference of π. When the frequency of the incident field is around the resonant frequency of the cavity, the leakage field has a comparable amplitude with the promptly reflected field, which results in significant destructive interference and ultimately leaves a notch in the spectrum space. **(B)** Illustration of the non-resonant and resonant coherent sum of the promptly reflected field and the leakage field.

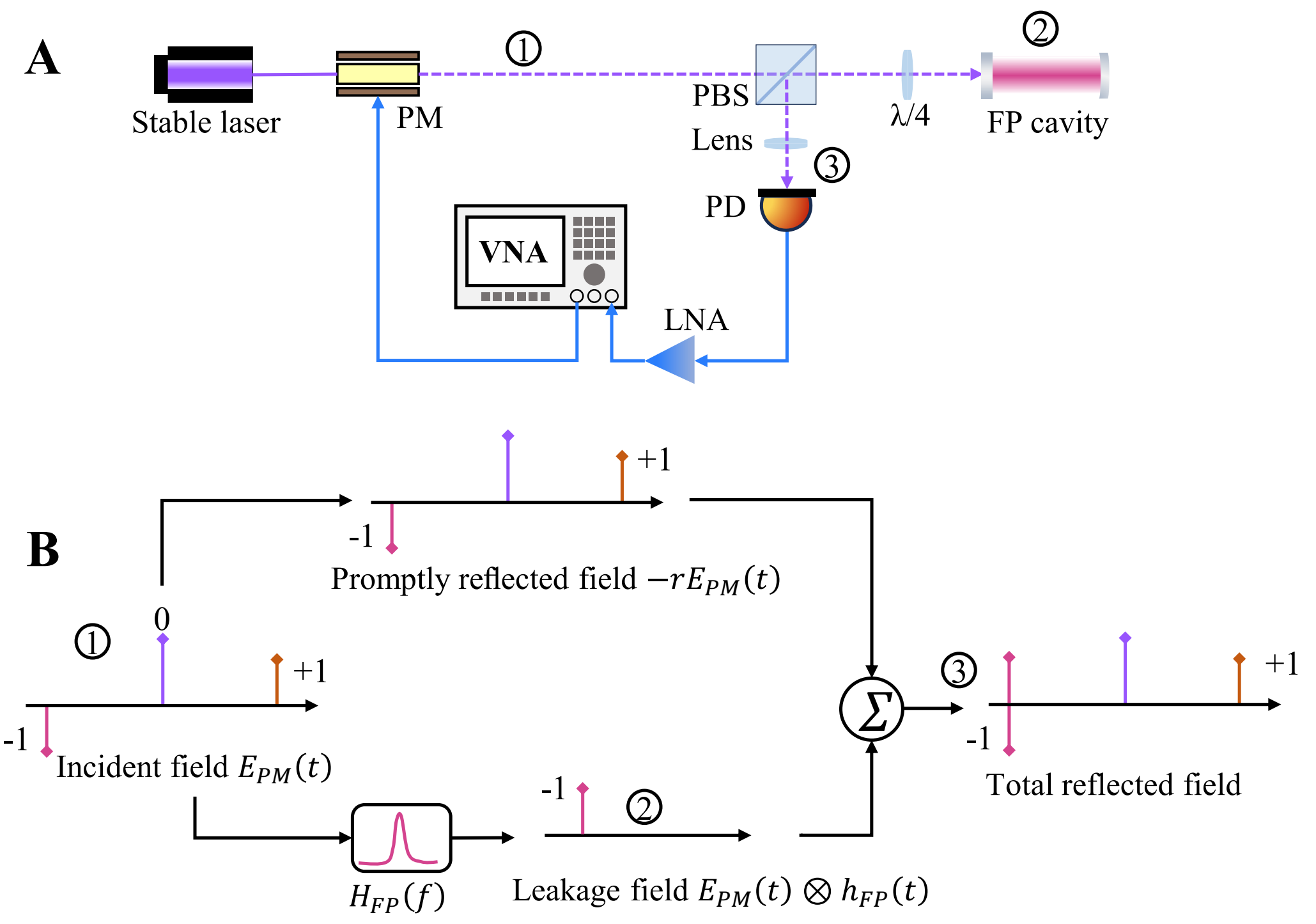

**Fig. S2. Measurement of the FP cavity response based on PM-IM-based MPF. (A)** Experimental setup of measuring the equivalent RF BPF based on PM-IM conversion. **(B)** The spectra at the corresponding location of (A). VNA: vector network analyzer; LNA: low-noise amplifier; PBS: polarization beam splitter.

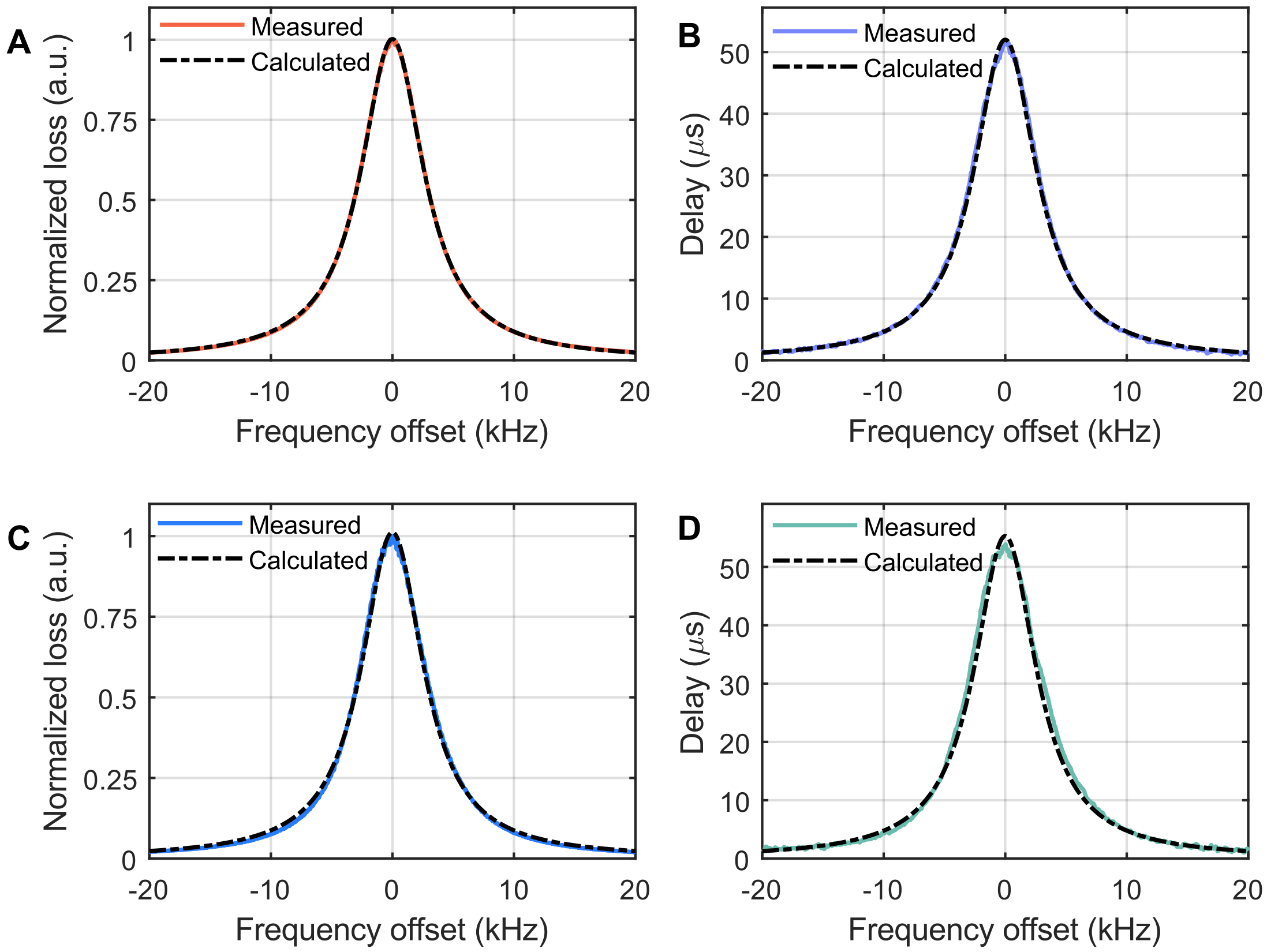

**Fig. S3. Measurements and calculations of RF BPFs based on the PM-IM conversion using the FP cavities. (A)** The normalized power loss and **(B)** the delay of the equivalent RF BPF based on FP cavity 1. **(C)** The normalized loss and **(D)** the delay of the equivalent RF BPF based on FP cavity 2. The bandwidth of the RF BPFs, as well as the linewidth of the corresponding FP cavities, are 6.26 kHz and 6.15 kHz, respectively. The experimental measurements agree with the theoretical calculations.

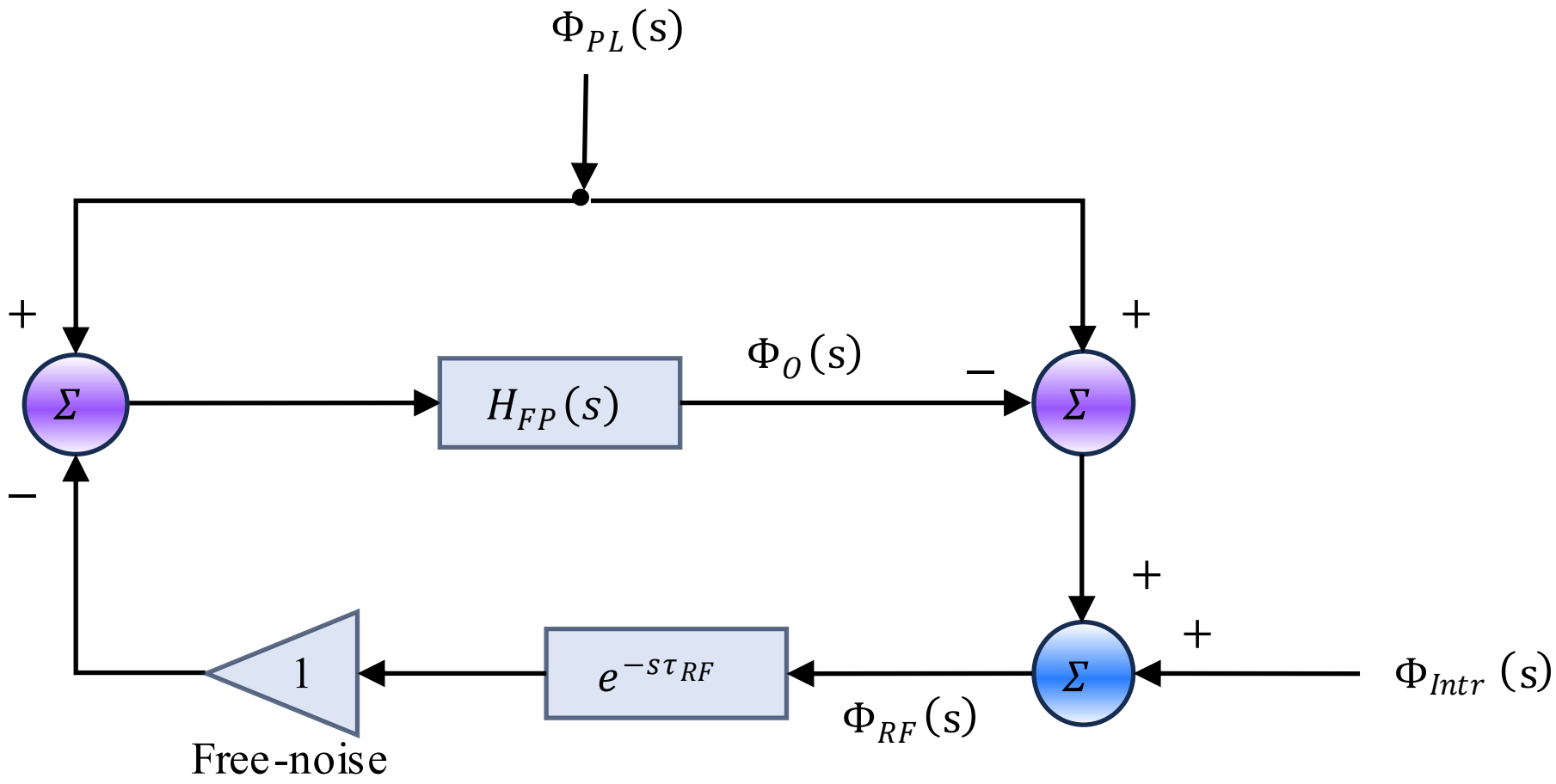

**Fig. S4. Input-output phase-noise model of the free-running OEO.** $\tau_{RF}$ is the delay in the optoelectronic link excluding the FP cavity.

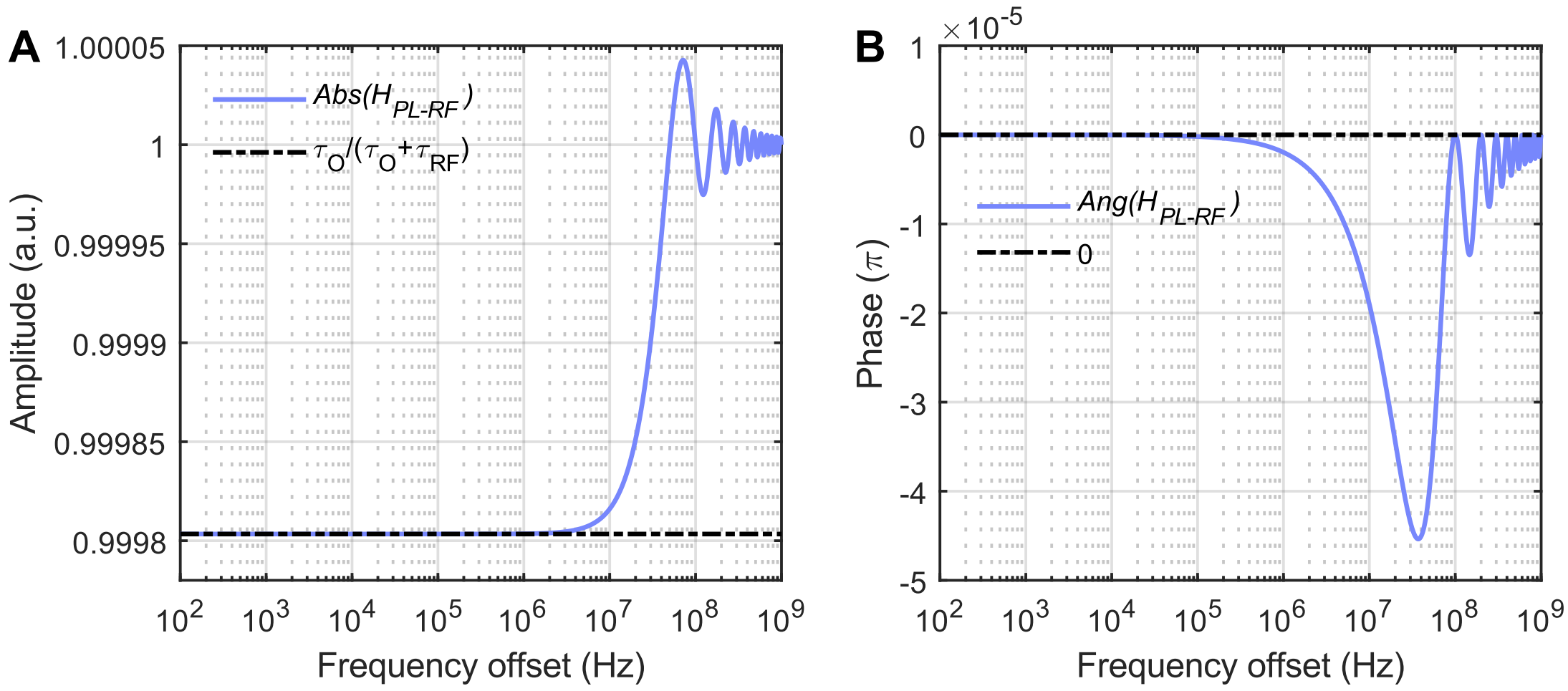

**Fig. S5. The calculation and approximation of the phase noise transfer function from the PL phase noise to the RF. (A)** The amplitude response. **(B)** The phase response.

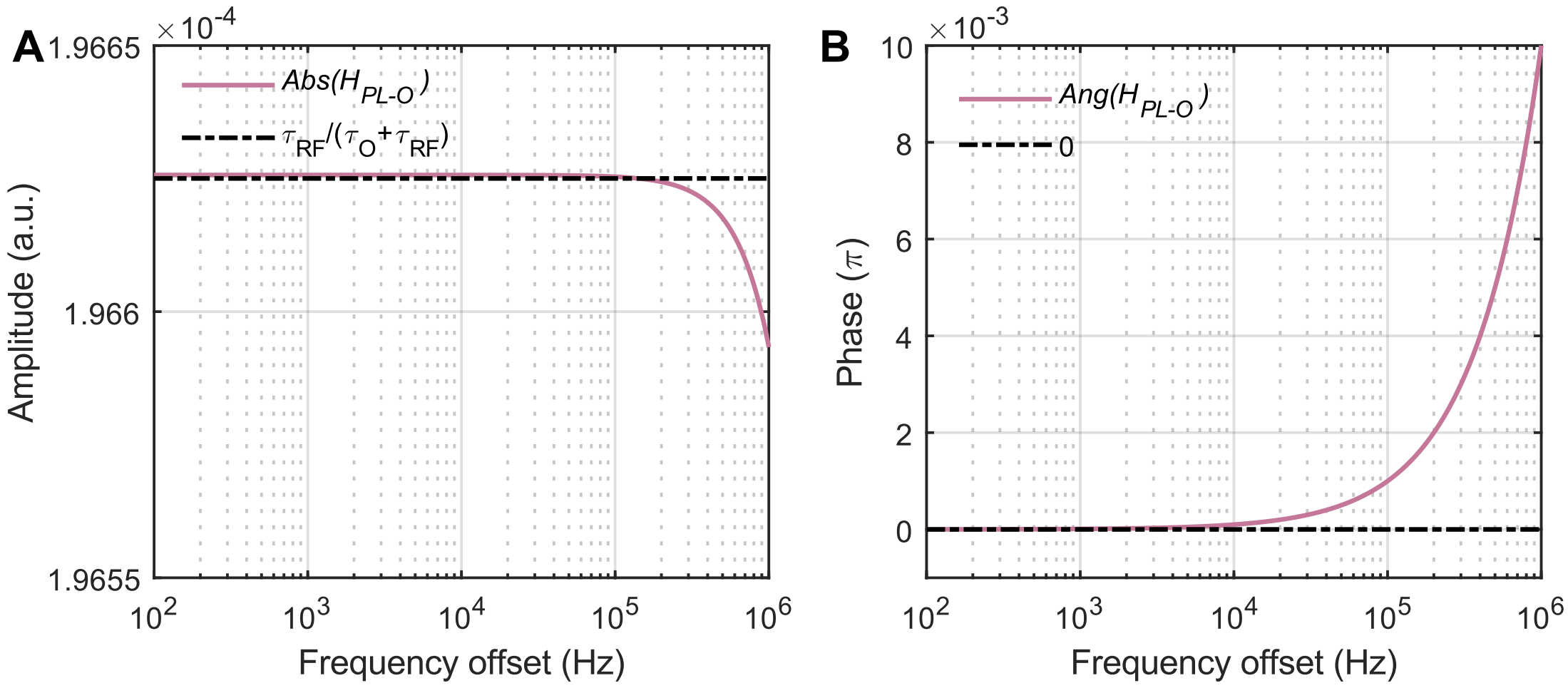

**Fig. S6. The calculation and approximation of the transfer function from the PL phase noise to the optical oscillation. (A)** The amplitude response. **(B)** The phase response.

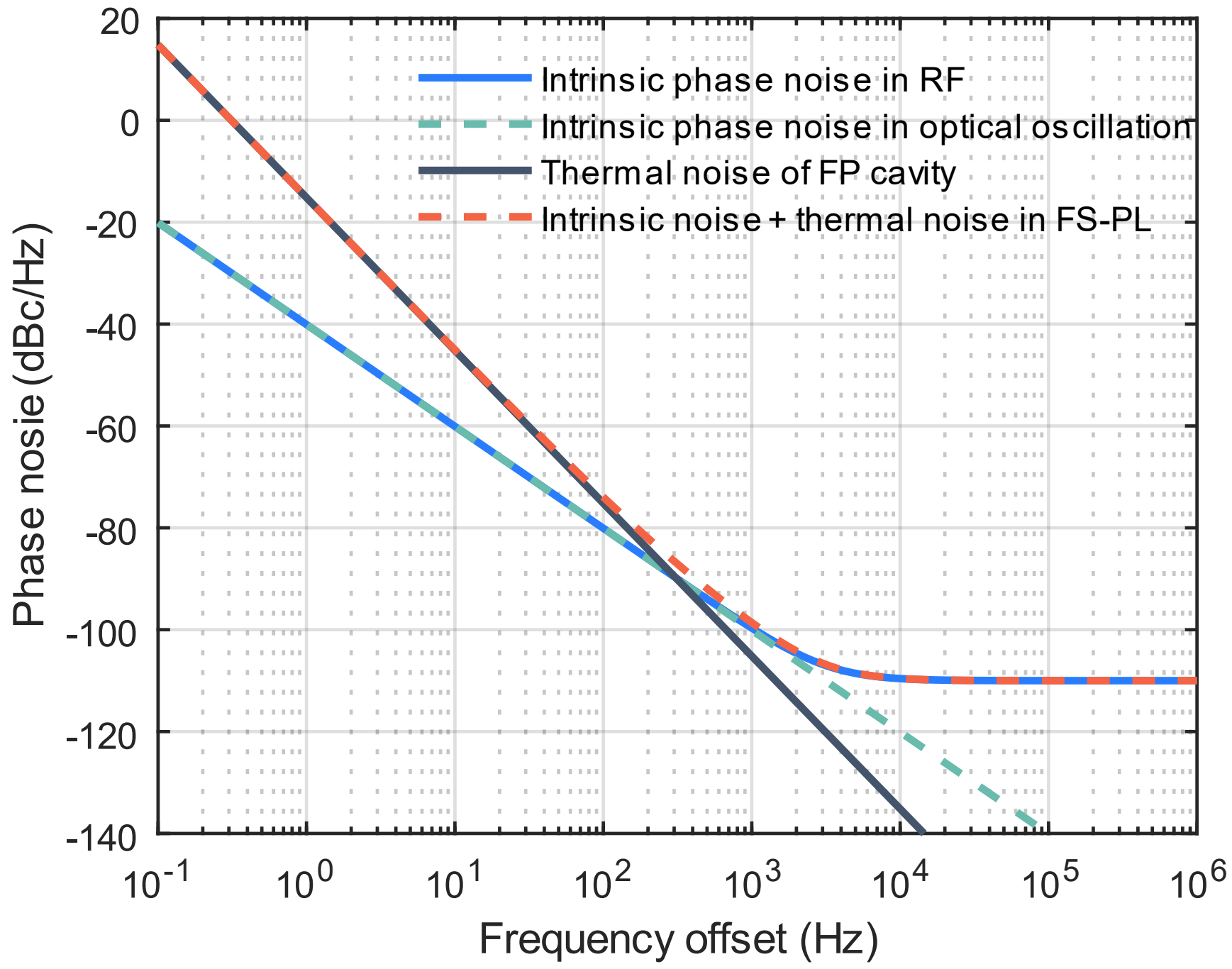

**Fig. S7. The phase noise introduced by intrinsic noise in the active optoelectronic link and thermal noise of the FP cavity.** At low-frequency offsets, the noise performance limit is determined by the thermal noise of the FP cavity, while at high-frequency offsets, it is dominated by the intrinsic noise.

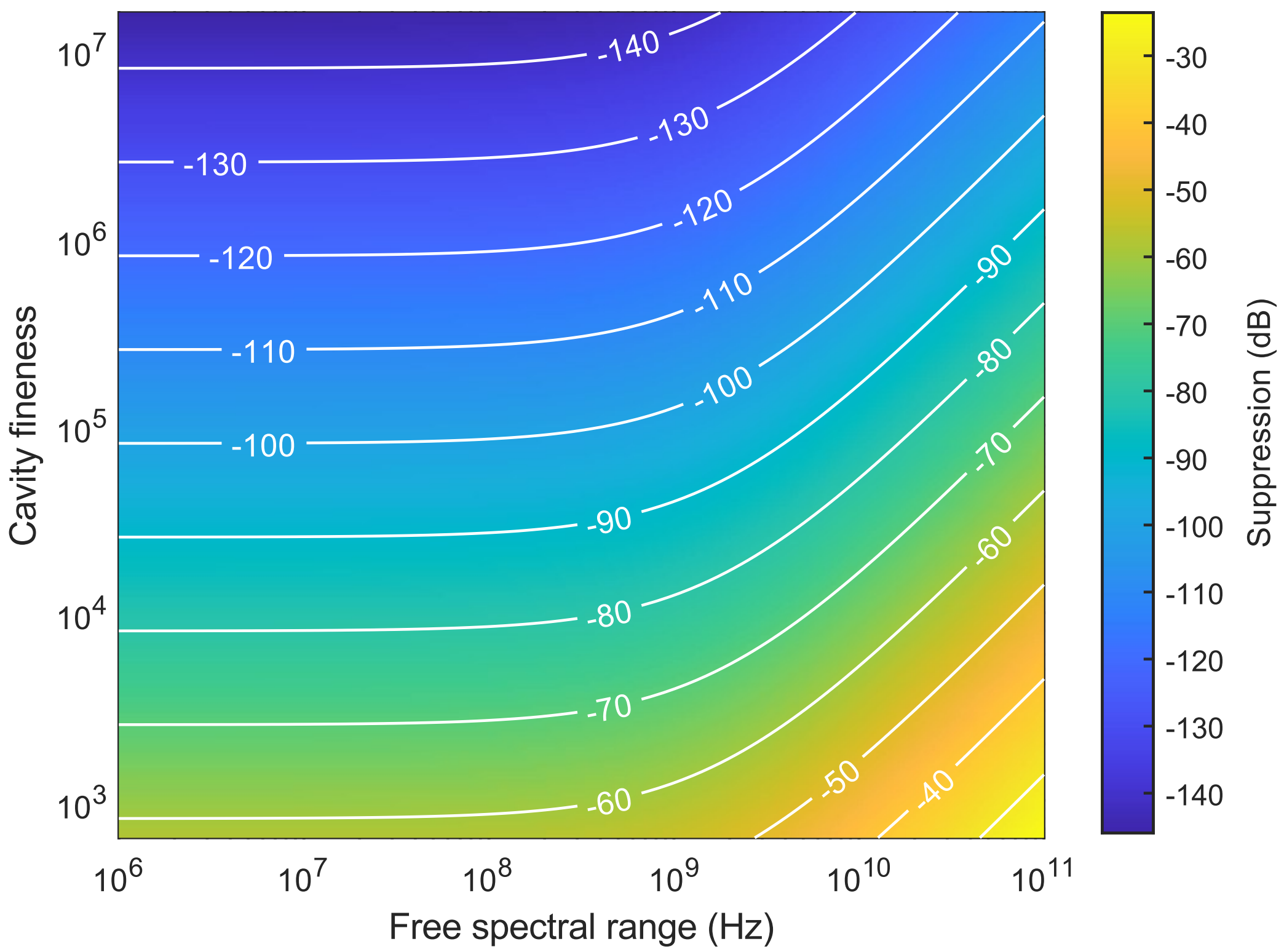

**Fig. S8. The phase noise suppression in the optical oscillation as a function of the cavity FSR and the fineness.** The flat contour line on the left indicates that the suppression has reached a limit determined by the delay introduced by the LPF. On the other hand, the sloping contour line on the right indicates that the delay in the optoelectronic link is primarily contributed by the delay from the optoelectronic components and connections between these components.

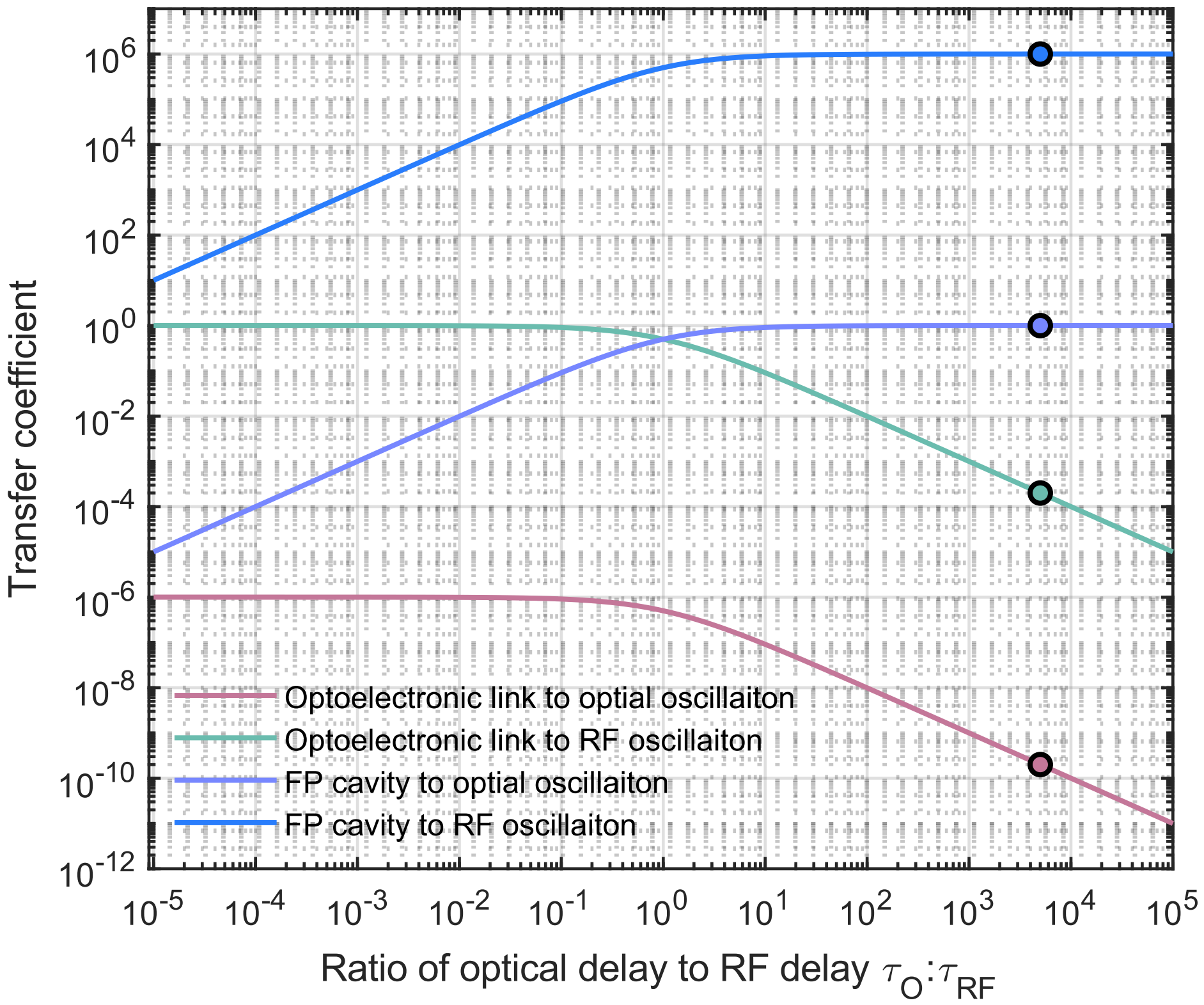

**Fig. S9. The transfer coefficients of the delay instability to the frequency instability as a function of the delay ratio.** The markers correspond to the parameters used in our experiment.

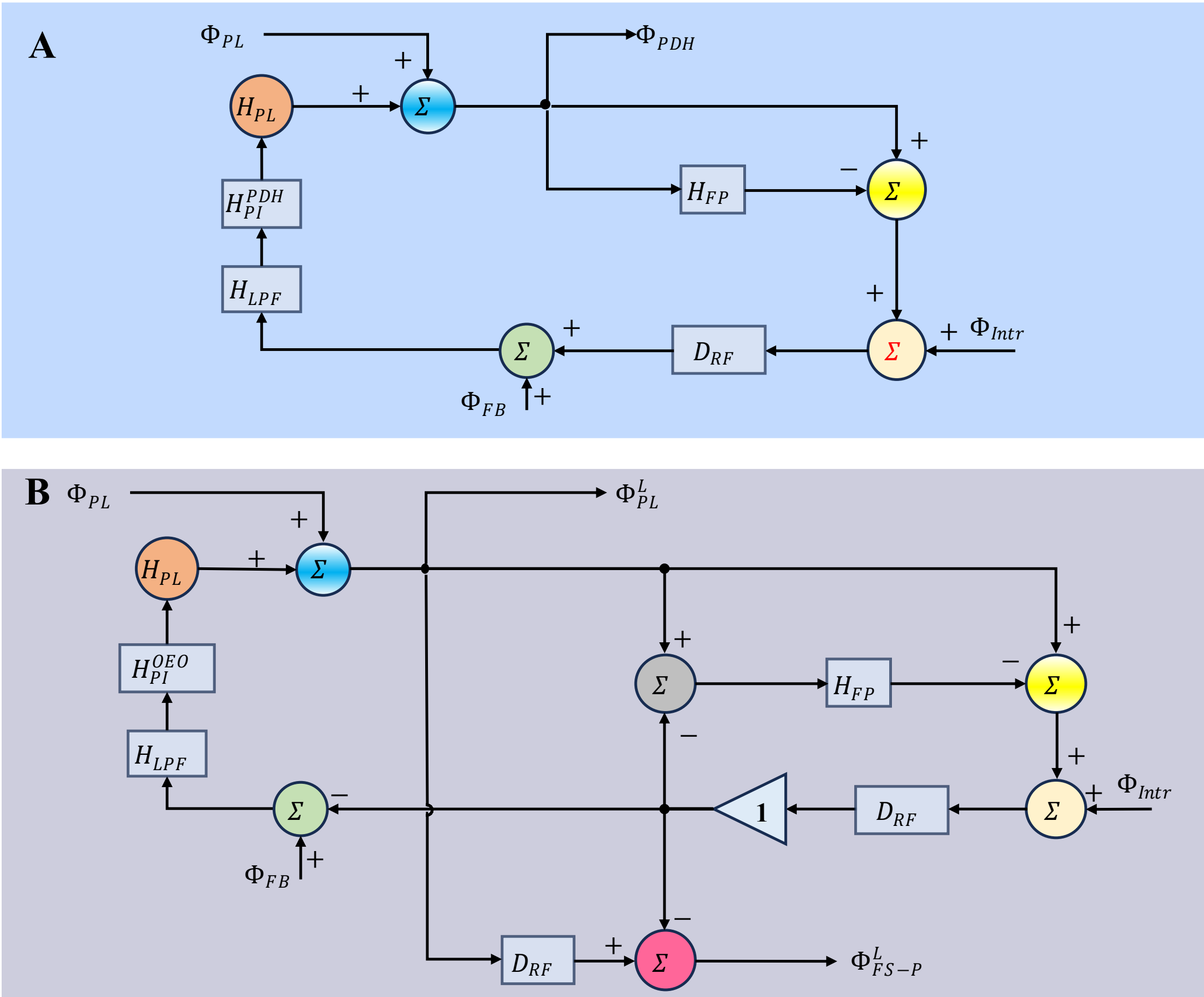

**Fig. S10. The input-output phase noise of (A)** the PDH scheme, and **(B)** the OEO-locking system. The parameters and the transfer functions are listed in Table S1 and Table S2. For simplicity, the gain of the phase detector is assumed to be 1 V/rad.

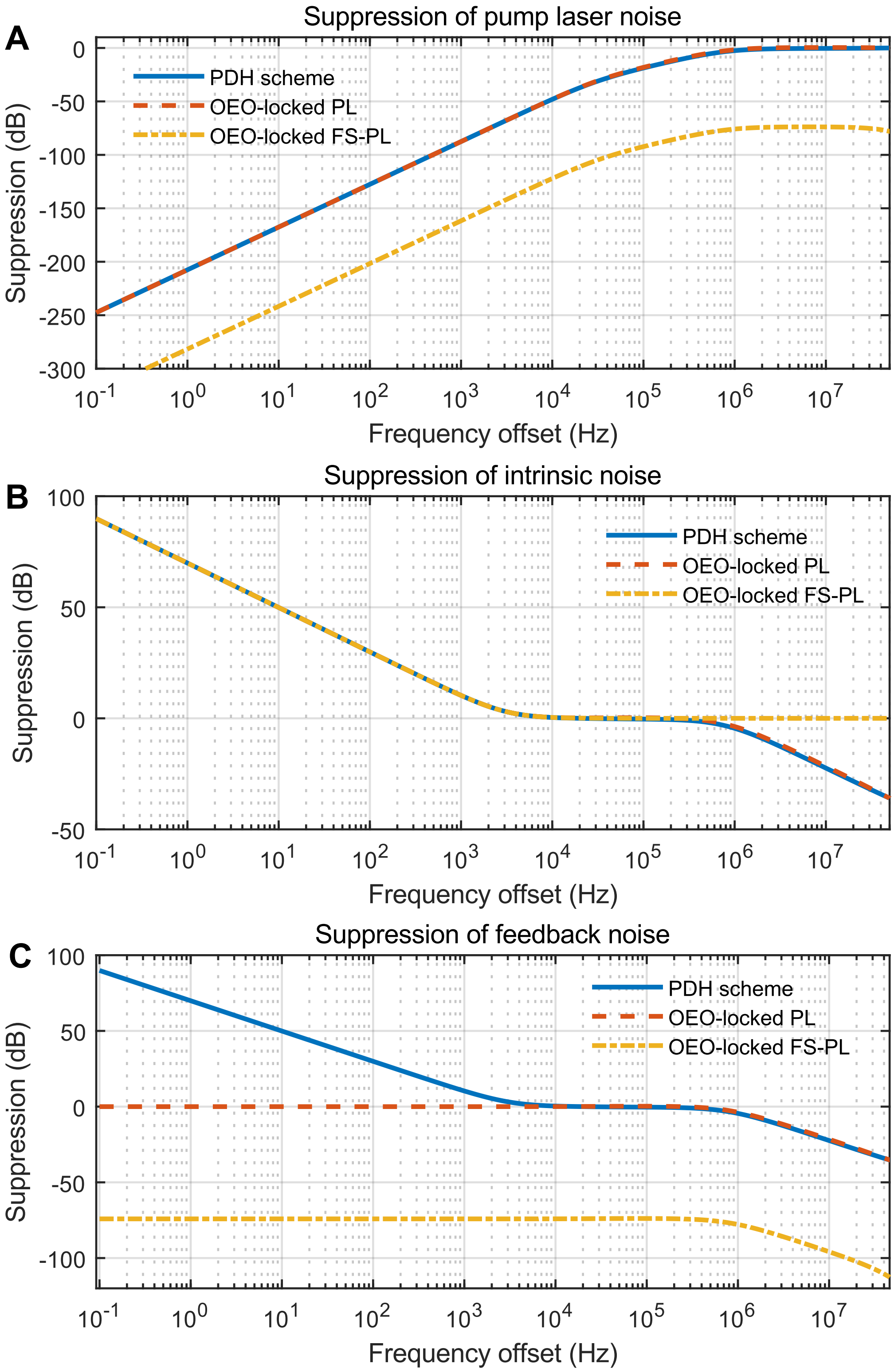

**Fig. S11. The phase noise suppressions in different systems. (A)** Suppression of PL phase noise. **(B)** Suppression of intrinsic noise. **(C)** Suppression of feedback link noise.

**Table S1. Input and output noise in PDH and OEO-locking systems**

| Symbol | Parameter meaning |
|---|---|
| $\Phi_{PL}(s)$ | Phase noise of free-running PL |
| $\Phi_{Intr}(s)$ | Intrinsic noise introduced from error extraction link |
| $\Phi_{FB}(s)$ | Noise introduced from feedback link |
| $\Phi_{PDH}(s)$ | Phase noise of PDH-locked laser |
| $\Phi_{PL}^{L}(s)$ | Phase noise of the OEO-locked PL |
| $\Phi_{FS-PL}^{L}(s)$ | Phase noise of OEO-locked FS-PL |

**Table S2. Transfer functions in PDH and OEO-locking systems**

| Symbol | Transfer function of | Mathematical expression | Parameter meaning | Value |
|---|---|---|---|---|
| $H_{FP}(s)$ | FP cavity | $\frac{Te^{-\frac{s}{f_{FSR}}}}{1-Re^{-\frac{s}{f_{FSR}}}}$ | $\Delta f_{BW}$: FP bandwidth; $f_{FSR}$: FP FSR. | $\Delta f_{BW}=6.26\ \text{kHz}$; $f_{FSR}=3\ \text{GHz}$ |
| $H_{LPF}(s)$ | First-order LPF | $\frac{2\pi f_c}{2\pi f_c+s}$ | $f_c$: Cut-off frequency. | $f_c=1\ \text{MHz}$ |
| $H_{PI}^{OEO}(s)$ | PI controller in OEO-locking system | $\frac{2\pi f_{PI}}{s}+1$ | $f_{PI}$: Integration bandwidth. | $f_{PI}=3\times10^4\ \text{Hz}$ |
| $H_{PI}^{PDH}(s)$ | PI controller in PDH system | $\left(\frac{\pi\Delta f_{BW}}{s}+1\right)H_{PI}^{OEO}(s)$ | / | $f_{PI}=3\times10^4\ \text{Hz}$ |
| $H_{PL}(s)$ | PL | $\frac{K_{PL}}{s}$ | $K_{PL}$: Sensitivity of the PL. | $K_{PL}=5\times10^6$ rad/V |
| $D_{RF}(s)$ | Optoelectronic link | $e^{-s\tau_{RF}}$ | $\tau_{RF}$: Delay in optoelectronic link | $\tau_{RF}=10\ \text{ns}$ |